\documentclass[preprint,times,tighten]{aastex}
\pdfoutput=1 
\usepackage{amsmath,amstext}

\usepackage{aas_macros}
\usepackage{rotating}
\usepackage{tabularx}
\usepackage{natbib}
\slugcomment{Accepted for publication in the Astronomical Journal}

\shorttitle{Discovery and Vetting of Exoplanets}
\shortauthors{Kostov et al.}

\newcommand{\alltargets}{$772~$}
\newcommand{\PCs}{$676~$}
\newcommand{\FPs}{$96~$}
\newcommand{\FPsSigSec}{$49~$}
\newcommand{\FPsCO}{$25~$}
\newcommand{\FPsOOTMOD}{$17~$}
\newcommand{\FPsLCMOD}{$8~$}
\newcommand{\FPsOE}{$6~$}
\newcommand{\FPsMoreThanOne}{$9~$}
\newcommand{\FPsVSHAPE}{$6~$}
\newcommand{\FPsVDE}{$48~$}
\newcommand{\NewFPs}{$60~$}
\newcommand{\OldFPs}{$36~$}
\newcommand{\OldFPsnotFound}{$25~$}
\newcommand{\OldFPsnotFoundnotFlagged}{$11~$}
\newcommand{\OldFPsnotFoundbutFlagged}{$14~$}
\newcommand{\NExScIConfirmed}{$276~$}
\newcommand{\NExScIFalse}{$61~$}
\newcommand{\NExScICandidates}{$435~$}

\begin{document}

\title{Discovery and Vetting of Exoplanets I: Benchmarking K2 Vetting Tools}
\author{
Veselin B. Kostov\altaffilmark{1,2},
Susan E. Mullally\altaffilmark{3,4},
Elisa V. Quintana\altaffilmark{1},
Jeffrey L. Coughlin\altaffilmark{2},
Fergal Mullally\altaffilmark{2},
Thomas Barclay\altaffilmark{1,6},
Knicole D. Col\'{o}n\altaffilmark{1},
Joshua E. Schlieder\altaffilmark{1},
Geert Barentsen\altaffilmark{7,8},
Christopher J. Burke\altaffilmark{9}
}

\altaffiltext{1}{NASA Goddard Space Flight Center, 8800 Greenbelt Road, MD 20771}
\altaffiltext{2}{SETI Institute, 189 Bernardo Ave, Suite 200, Mountain View, CA 94043, USA}
\altaffiltext{3}{a.k.a. Susan E. Thompson}
\altaffiltext{4}{Space Telescope Science Institute, 3700 San Martin Dr., Baltimore MD 21212}
\altaffiltext{5}{Orbital Insight, 3000 El Camino Real bldg 2 floor 6, Palo Alto, CA 943}
\altaffiltext{6}{University of Maryland, Baltimore County}
\altaffiltext{7}{NASA Ames Research Center, M/S 244-30, Moffett Field, CA 94035}
\altaffiltext{8}{Bay Area Environmental Research Institute, 625 2nd St. Ste 209, Petaluma, CA 94952}
\altaffiltext{9}{MIT Kavli Institute for Astrophysics and Space Research, 70 Vassar St., Cambridge, MA 02139, USA}

\begin{abstract}
We have adapted the algorithmic tools developed during the Kepler mission to vet the quality of transit-like signals for use on the K2 mission data. Using the four sets of publicly-available lightcurves on MAST, we produced a uniformly-vetted catalog of  \alltargets transiting planet candidates from K2 as listed at the NASA Exoplanet archive in the K2 Table of Candidates. Our analysis marks \PCs of these as planet candidates and \FPs as false positives. All confirmed planets pass our vetting tests. \NewFPs of our false positives are new identifications -- effectively doubling the overall number of astrophysical signals mimicking planetary transits in K2 data. Most of the targets listed as false positives in our catalog either show prominent secondary eclipses, transit depths suggesting a stellar companion instead of a planet, or significant photocenter shifts during transit. We packaged our tools into the open-source, automated vetting pipeline DAVE (Discovery and Vetting of Exoplanets) designed to streamline follow-up efforts by reducing the time and resources wasted observing targets that are likely false positives. DAVE will also be a valuable tool for analyzing planet candidates from NASA's TESS mission, where several guest-investigator programs will provide independent lightcurve sets---and likely many more from the community. We are currently testing DAVE on recently-released TESS planet candidates and will present our results in a follow-up paper.

\end{abstract}

\keywords{planets and satellites; techniques:photometric}
\maketitle

\section{Introduction}

From the numerous ground and space-based studies that have detected exoplanets, it has become evident over the past 20+ years that planetary systems are not only common, but diverse in nature as well (e.g. Borucki et al. 2011; Burke et al. 2015, Thompson et al. 2018). NASA's Kepler mission has shown that the occurrence rate for small planets is high, with almost every low-mass star expected to host at least one small ${\rm (R<4 R_\oplus)}$ planet (e.g.~Dressing \& Charbonneau 2015). While missions like Kepler (Borucki et al. 2010), CoRoT, and various ground-based photometric and radial-velocity surveys have successfully expanded our knowledge of exoplanets, there are still regions of parameter space that are largely unexplored. These include planets around nearby stars and small planets around bright stars; systems that K2---the repurposed Kepler mission---is well-suited to explore. The major advantage of observing both bright and nearby stars is that planets can be more than detected---they can be characterized in detail. Such planets can be studied by a) precise Doppler spectroscopy to get their masses and densities; b) both emission and transmission spectroscopy to characterize their atmospheric properties; c) high-contrast direct imaging to search for longer period planets; and d) asteroseismology to determine precise stellar properties---essential for precise planetary radii, masses and equilibrium temperatures.

NASA's K2 mission has been tremendously successful at finding planets, as demonstrated by a number of published catalogs containing hundreds of planet candidates, with many of them validated or confirmed (e.g.~Montet et al. 2015, Adams et al. 2016, Barros et al. 2016, Crossfield et al. 2016, Pope et al. 2016, Vanderburg et al. 2016, Cabrera et al. 2017, Dressing et al. 2017, Rizzuto et al. 2017, Shporer et al. 2017, Mayo et al. 2018, Livingston et al. 2018). Some of these planets are as small as the Earth and transit bright, nearby stars. These targets will be particularly well-suited for follow-up observations with the James Webb Space Telescope (JWST) with the goal to study their atmospheric composition and density. Such observations will yield a better understanding of the difference between rocky and gaseous planets, particularly how composition varies as a function of planet radius. 

A major challenge for transit surveys like K2 targeting a large number of stars is to distinguish false-positive (eclipses/transits not due to planets) and false-alarm signals (signals unrelated to any eclipsing/transiting astrophysical system) from real transit events. An additional complication associated with K2 is the significant systematics due to spacecraft pointing drift (Howell et al. 2014), which introduces a significant difficulty in detecting high-quality planet candidates. Another layer of difficulty is introduced when there are multiple independent datasets per target---each employing a different approach to systematics correction on a star-by-star basis. This is indeed the case for K2 where, as described below, four different teams have produced publicly-available lightcurve sets\footnote{And will be the case for TESS as well.}. Each detected planet candidate needs to undergo proper vetting to eliminate instrumental artifacts, non-transiting variable stars, eclipsing binaries, and contamination from variable objects other than the target star. In the absence of radial velocity measurements to confirm the planetary nature of a transit-containing lightcurve, one has to rely on additional methods to distinguish between a bona-fide planet and a false positive. To tackle such obstacles, a number of methods have been developed for the Kepler mission, each tailored to a particular source of potential false positives, and over successive catalogs of planet candidates as the data and sources of false alarms and false positives were better understood (e.g. Borucki et al. 2010; Batalha et al. 2011; Coughlin et al. 2016; Mullally et al. 2015; Rowe et al. 2015; Thompson et al. 2015, 2018). 

Several automated vetting codes have been developed and applied to Kepler data. For example, the Autovetter (Catanzarite 2015) uses machine-learning random forest decision tree to classify Threshold Crossing Events (TCEs) as planet candidates, false positives, or non-transiting phemonenon. Robovetter (Coughlin et al. 2016) utilizes specifically designed metrics that mimic the decision process of human vetting to assign a TCE as a planet candidate or a false positive. Vespa (Morton 2012) uses a probabilistic algorithm designed to determine whether a transit-like signal is statistically likely to be caused by a background eclipsing binary. Astronet (Shallue \& Vanderburg 2018) and Exonet (Ansdell et al. 2018) are also machine-learning tools that use deep learning to classify transit signals in Kepler data, including analysis of centroid time-series and scientific domain knowledge.

Here, we present a new pipeline designed for the Discovery And Vetting of Exoplanets (DAVE) from K2. DAVE implements vetting tools used for the Kepler mission (e.g.~Coughlin et al. 2014; Thompson et al. 2015, 2018;  Mullaly 2017) to produce a uniformly-vetted catalog of K2 planet candidates. The pipeline adapts these tools to K2 data and focuses on applying robotic vetting techniques, formulated as part of the prime Kepler mission, to K2 data. We highlight these techniques and the types of false positives they eliminate, and present a robust catalog of uniformly-vetted planet candidates and false positives. We have reclassified \NewFPs targets previously listed as planet candidates as false positives, and two targets previously listed as a false positive as planet candidates (EPIC 211970234.01 and EPIC 212572452.01). DAVE's key benefit to the community is the potential to reduce the effort expended following up on false positives because we can identify them with Kepler data alone. The DAVE pipeline, including the vetting tools, is publicly available at https://github.com/exoplanetvetting/DAVE. 

We note that while K2 and Kepler data are based on the same instrument, the two sets have different characteristics and systematics (e.g.~K2 data does not span multiple quarters, has a significant roll motion). Thus vetting tools trained for Kepler data, such as Autovetter, Astronet and Exonet, may not be optimal for application to K2 data. 
Compared to machine-learning algorithms, DAVE has the advantage of well-defined individual tests---thoroughly scrutinized across multiple Kepler planet catalogs---that provide specific reasons for failing a particular planet candidate. With that said, while we have ported over some parts of the Robovetter into DAVE (e.g.~Model-shift), DAVE does not (yet) have the capability to measure the completeness and reliability of the vetting process, or the validation power of Vespa. 

\subsection{Different lightcurves, different vetting}
\label{sec:LCs}

Throughout this work, when available, we use four different sets of K2 lightcurves in our DAVE analyses. The four lightcurve reductions used are: Aigrain et al. 2016 (AGP), Luger et al. 2016 (EVEREST), Vanderburg et al. 2014 (k2sff), and the Kepler/K2 program office processed lightcurves (PDC).

The Aigrain et al. (2016) AGP lightcurves are the result of three-component Gaussian process used to model the observed stellar flux and remove systematics. The three components include one with a dependence on pixel position, one with a dependence on time, and a white noise component. The lightcurves from the Luger et al. (2016, 2018) EVEREST algorithm use a variant of the pixel-level decorrelation technique developed to correct systematics in Spitzer data (Deming et al. 2015). The pixel-level decorrelation procedure removes spacecraft pointing induced systematics and a Gaussian process is then used to model time dependent astrophysical variability. We used the EVEREST lightcurves as available on MAST, without masking-and-recomputing. The k2sff lightcurves use a "self-flat-fielding" (SFF, Vanderburg et al. 2014) method to remove photometric variability due to the imprecise pointing of K2. The SFF technique involves an iterative basis-spline fit to low-frequency variability and an iterative procedure to remove position dependent noise that depends on the arclength of the path a star follows on the detector. The K2 Mission creates Pre-Search Data Conditioning (PDC) ligtcurves which use a modified version of the Kepler processing algorithm (Smith et al. 2012). PDC removes systematic errors using a process where flux signatures that are similar across many stars on the same group of modules are removed by fitting basis vectors using a Bayesian approach.

The different approaches inherent to the systematics correction employed in each set of lightcurves leads to different lightcurve properties on a star-by-star basis. This in turn leads to the detection of both the same planet candidates with different properties (signal-to-noise ratio, transit depth etc.) and the inhomogeneous detection of different planet candidates depending on which lightcurves are searched, what planet search algorithm is used, and the details of subsequent human candidate vetting (e.g.~Crossfield et al. 2016, Crossfield et al. 2018). In this work we attempt to mitigate some of these biases by performing a uniform DAVE analysis on each lightcurve available for a given planet candidate and provide below lessons learned throughout this process. 

{\noindent This paper is organized as follows. In \S\ref{pipeline} we present our algorithm for generating detrended lightcurves, our search algorithm, and the different vetting metrics we applied. In \S\ref{sec:catalog} we outline the K2 target sample and describe our catalog of uniformly-vetted planet candidates. Finally, we draw our conclusions in \S\ref{end}.}

\section{DAVE Pipeline}
\label{pipeline}

The pipeline consists of several modules, each tailored to particular aspects of the vetting procedures we used. These are broadly split into two categories: A) photocenter analysis to rule out background eclipsing binaries; and B) flux time-series analysis to rule out odd-even differences, secondary eclipses, low signal-to-noise ratio (SNR) events, variability other than a transit, and size of the transiting object. The metrics these modules produce are described below.

\subsection{Vetting Metrics: Centroid Analysis}

Measuring the position of the photocenter of light during a planetary transit (or a stellar eclipse) is a powerful method to distinguish between a genuine occultation occurring in the target system and an unresolved background source aligned along the same line-of-sight (e.g. an eclipsing binary). Specifically, a strong indicator of a false positive is a photocenter shift away from the target's location on the detector during a transit. When applied to the original Kepler mission, this method was successful at identifying such false positives with a sub-pixel precision (Bryson et al. 2013). The roll motion of K2, however, changes the distribution of light on the detector from one cadence to the next, independent of any astrophysical variability. The centroid analysis used in DAVE extends the difference imaging technique described by Bryson et al. (2013), and is outlined in Christiansen et al (2017); we summarize it here for completeness. 

To compensate for the effects introduced by the roll motion of K2, we calculate the photocenter per in-transit cadence instead of per transit as follows. First, we find the in- and out-of-transit cadences having the same roll angle and separated by a single thruster firing event (to minimize the effects of velocity aberration on the roll axis). Next, we compute the photocenter offsets by fitting a pixel response function model (Bryson et al. 2010) to the out-of-transit and difference images. We note that the repeatability of the K2 roll motion is not perfect and thus in-transit cadences do not always have corresponding out-of-transit cadences at the same roll angle. Finally, assuming that the measured offsets follow a Gaussian distribution, we estimate the probability that these are statistically significant by averaging the centroid offset over all out-of-transit cadences and their corresponding difference cadences.

\begin{figure*}
\centering
\epsscale{0.85}
\plottwo{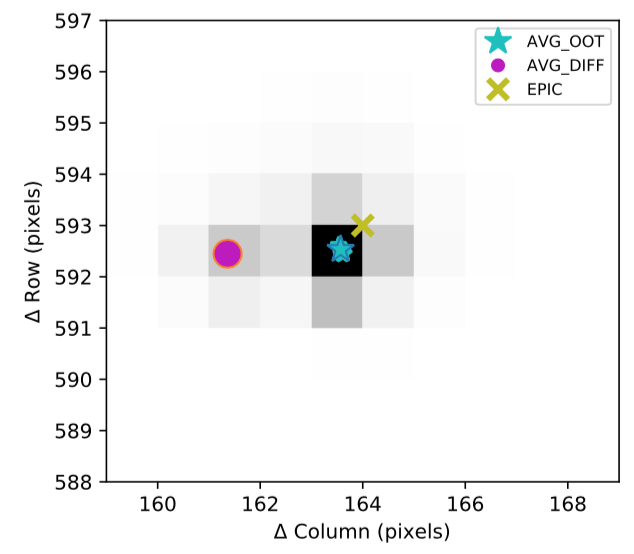}{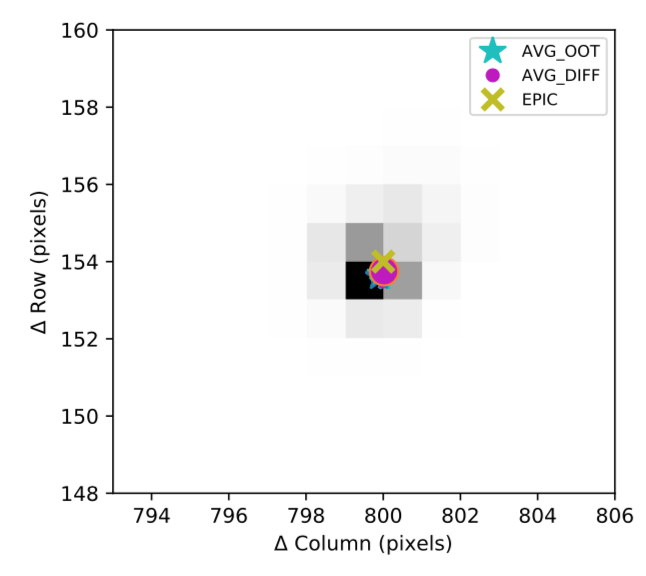}
\plottwo{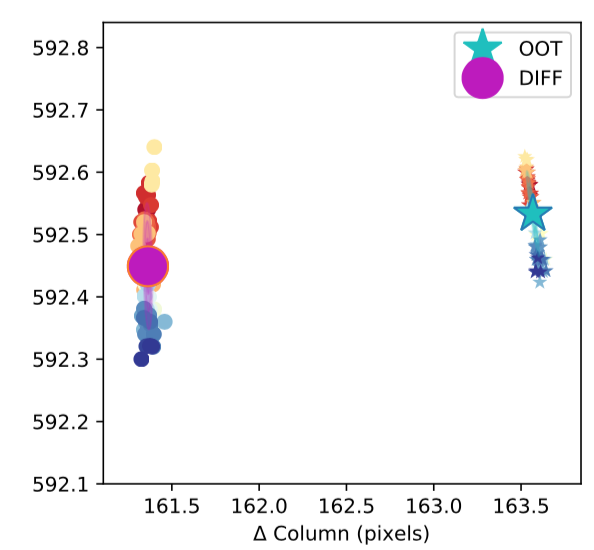}{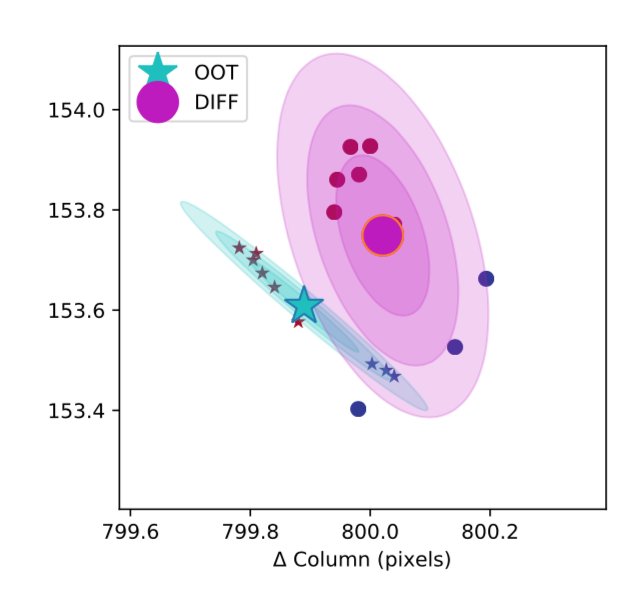}
\plottwo{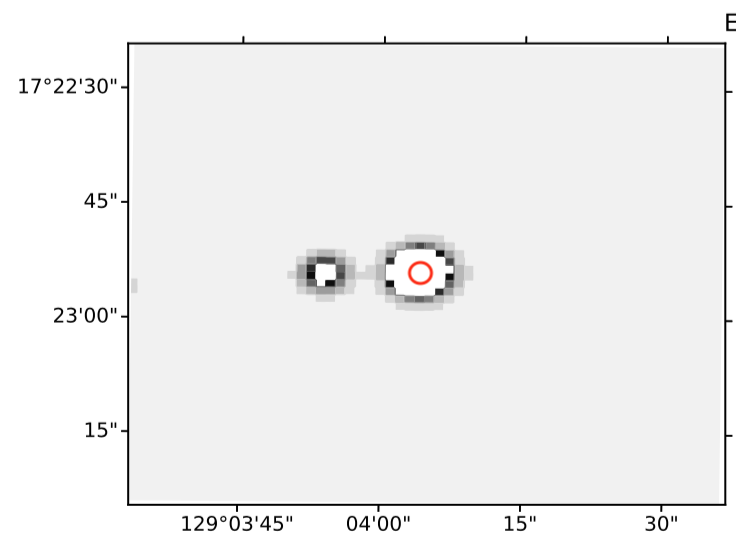}{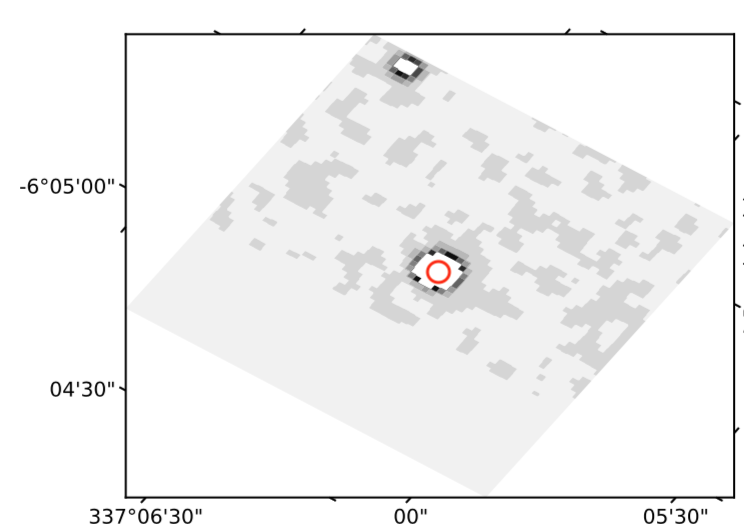}
\caption{An example of the centroid vetting module of DAVE, showing a target with a clear photocenter shift (left panels, EPIC 211804579) and another target with no significant photocenter shift (right panels, EPIC 206432863). Upper panels: The small circle symbols (magenta) represent individual difference image photocenter positions , and the large circle symbol represents the average position. The star symbols (cyan) represent the corresponding out-of-transit photocenter positions. The catalog position of the respective EPIC target is marked with a yellow X. Middle panels: same as upper panels but zoomed in to better show the individual photocenters, which are now also colored by the cadence number (from red to blue), along with the corresponding confidence intervals. Lower panels: Respective 1\arcmin x 1\arcmin 2MASS J-band images, with the K2 targets marked with a red circle. The source of the photocenter offset for EPIC 211804579 is clearly visible on the image as a field star to the east of the target star.
\label{fig:offset_}}
\end{figure*}

An example result from the centroid vetting module is shown in Fig. \ref{fig:offset_}, demonstrating one target with a clear photocenter shift (EPIC 211804579, listed as a false positive and a planetary candidate on both the NASA Exoplanet Archive and ExoFOP at the time of writing) and another with no significant shift (EPIC 206432863, false positive due to RV measurements, Shporer et al. 2017). 


We note that sometimes bright field stars inside the aperture of the target star may capture the out-of-transit PRF fit and thus result in a false centroid offset (e.g. KOI-1860, Bryson et al. 2013). Such cases are marked in our catalog as ``centroid offset spurious'' (COSp); if the separation between the target star and the field star(s) is at least one pixel, we use the lightkurve package (Vinícius et al. 2018) to extract custom lightcurves from apertures containing only the target star and only the field star(s). If the custom lightcurve demonstrates that the transit signal is coming from the target star, it is listed in our catalog as a planet candidate. Alternatively, if the field star is the source of the signal we flag the target as a false positive. 

\begin{figure*}
\centering
\plotone{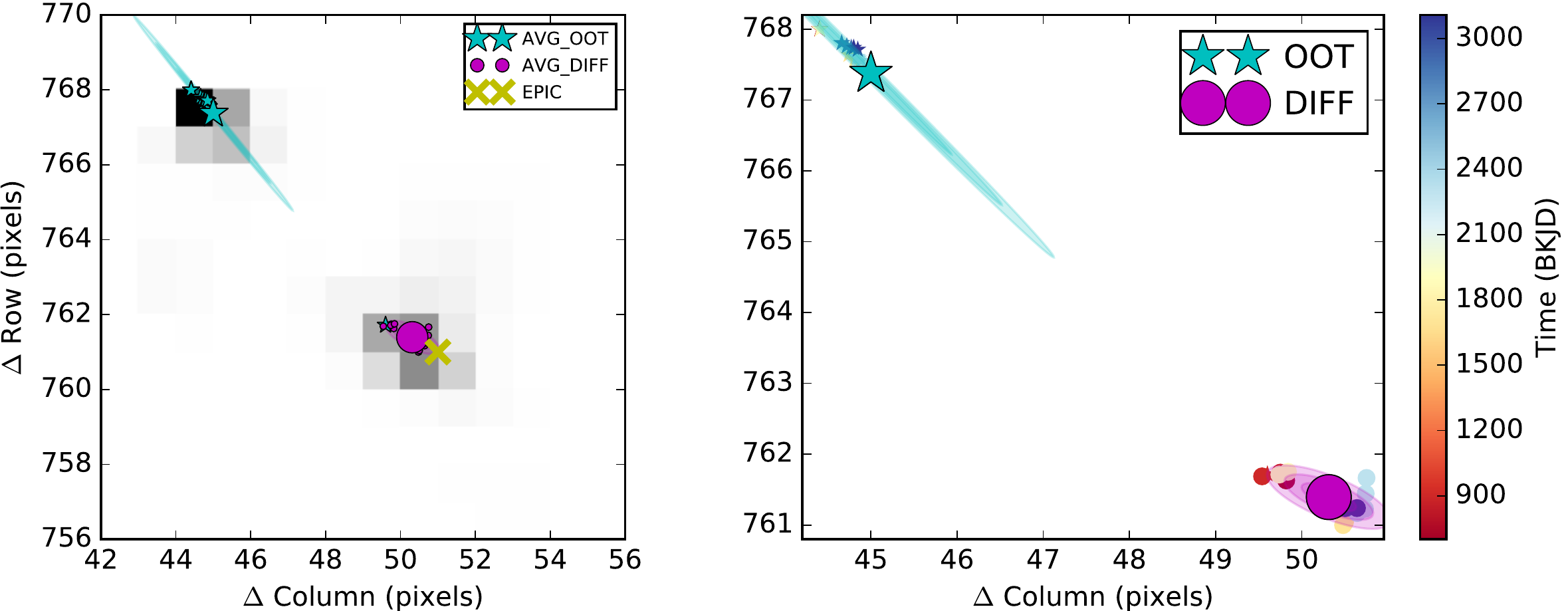}
\plotone{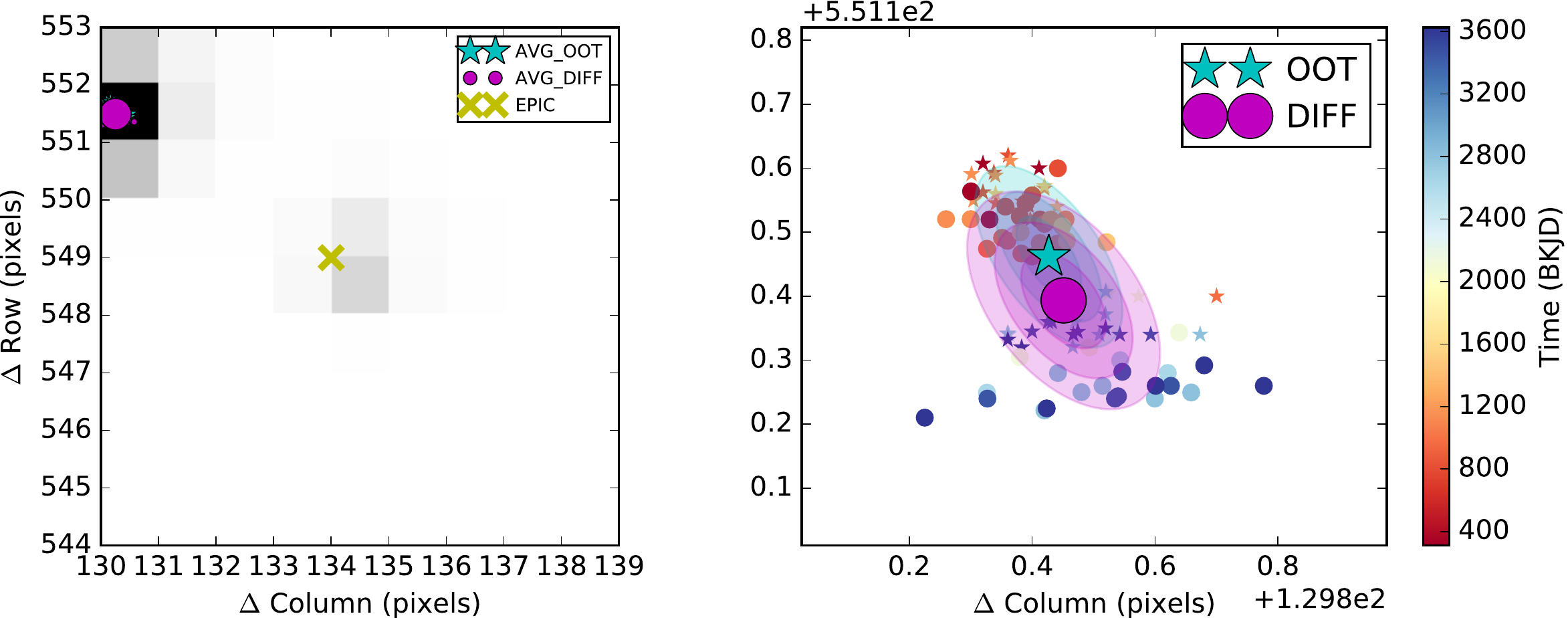}
\caption{Upper panels: An example of a measured spurious centroid offset for a planet candidate where the out-of-transit photocenter (cyan) is locked onto a bright field star but the different centroid (magenta) is locked onto the target star (EPIC 210957318) itself (yellow X), which is the source of the signal. Right panel is the same as the left panel but zoomed-in to better show the individual centroid measurements with their respective confidence intervals. Lower panels: An example of a measured spurious centroid offset for a false positive (EPIC 211808055), where both the out-of-transit and difference photocenters are locked onto the bright field star, which is the source of the signal. Custom lightcurves extracted for EPIC 210957318 and EPIC 211808055 are shown in Figures \ref{fig:COSp_PC_LK} and \ref{fig:COSp_FP_LK} respectively.  
\label{fig:COSp_}}
\end{figure*}

\begin{figure*}
\centering
\plottwo{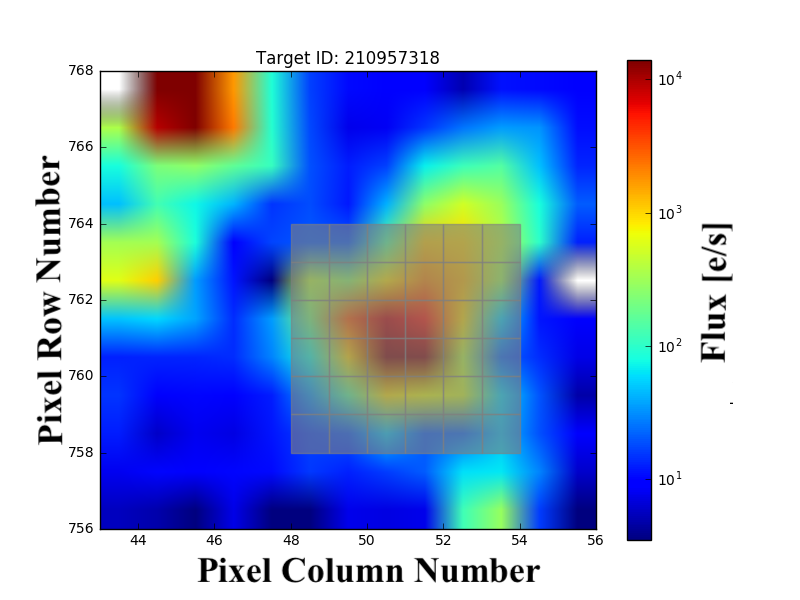}{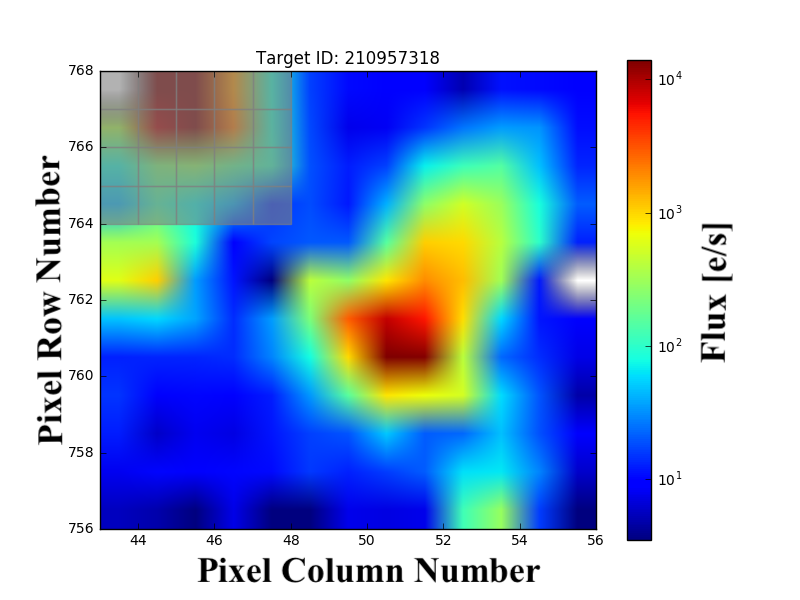}
\plottwo{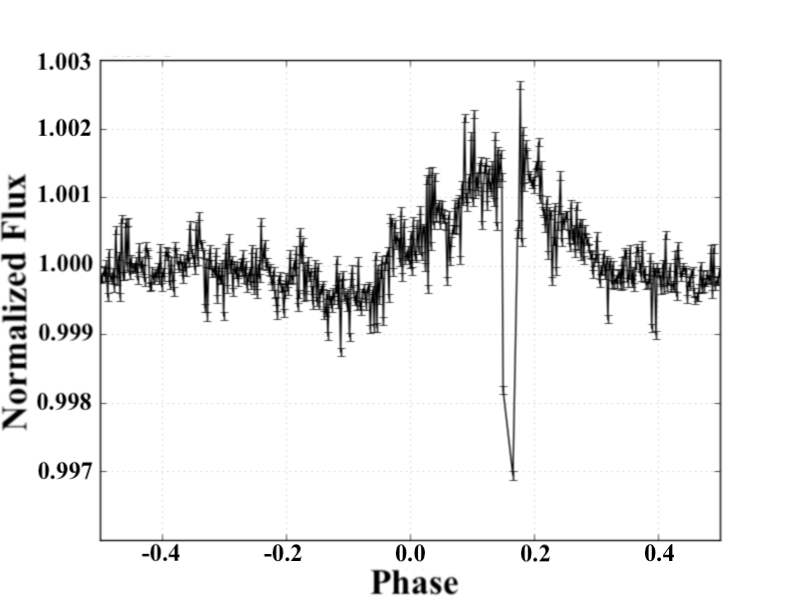}{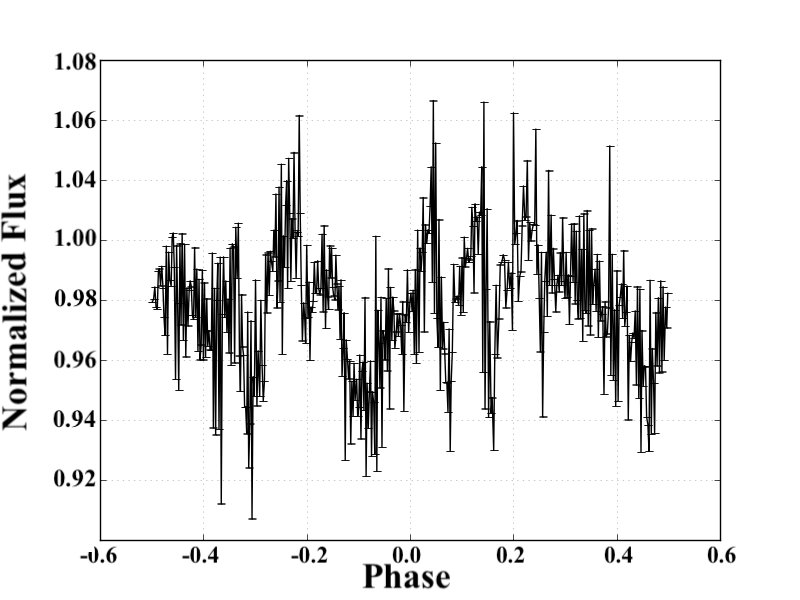}
\caption{Detailed analysis of the planet candidate EPIC 210957318 using lightkurve. Upper panels: Custom apertures centered on the target star (left) and on the field star (right). Lower panels: Extracted lightcurves for the target star (left) and field star (right), demonstrating that the source of the signal is the target star itself and the measured centroid offset in Figure \ref{fig:COSp_} (upper panels) is spurious.
\label{fig:COSp_PC_LK}}
\end{figure*}

\begin{figure*}
\centering
\plottwo{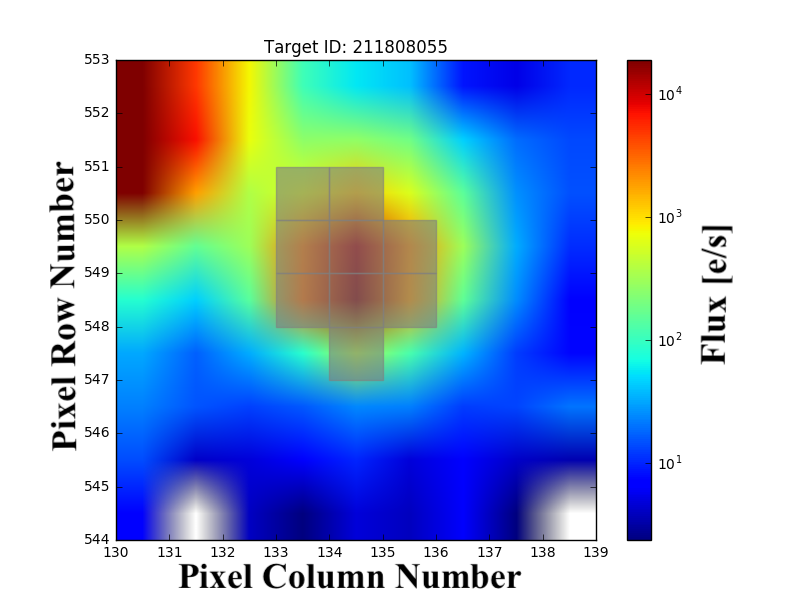}{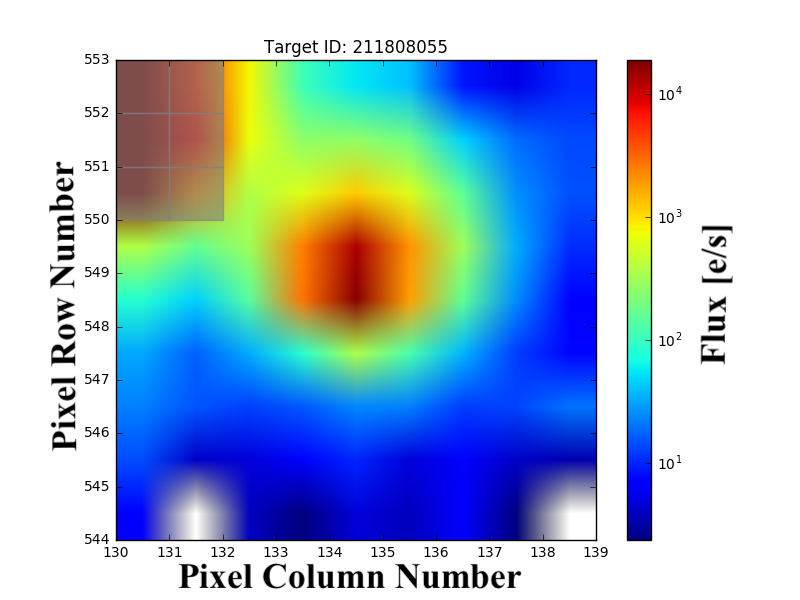}
\plottwo{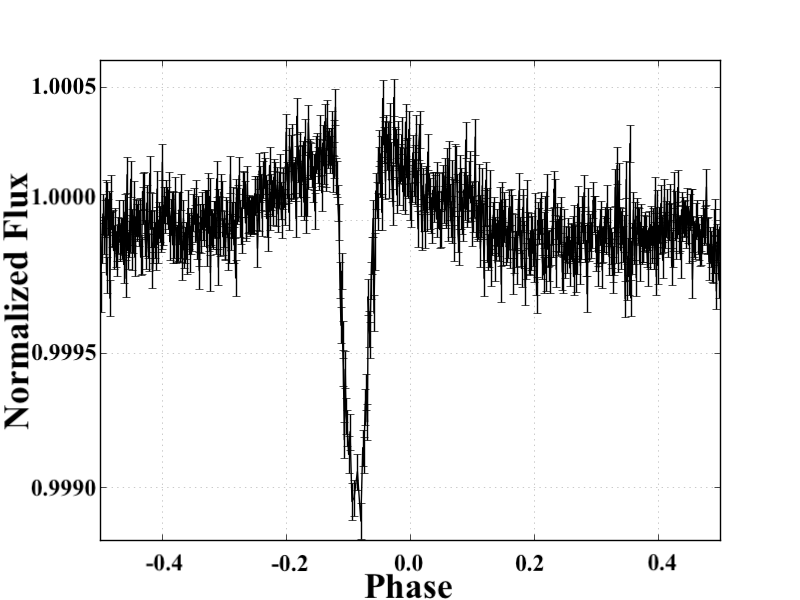}{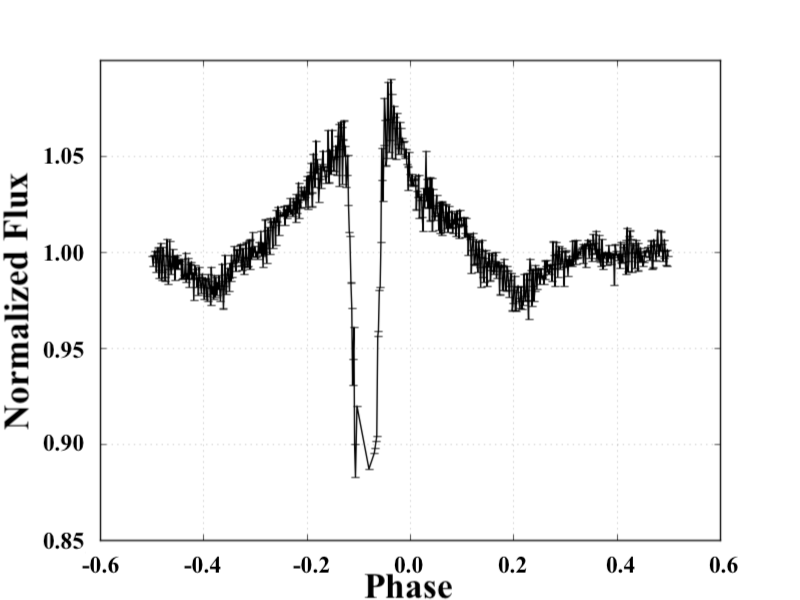}
\caption{Same as Figure \ref{fig:COSp_PC_LK} but for the false positive EPIC 211808055. Upper panels: Custom apertures centered on the target star (left) and on the field star (right). Lower panels: Extracted lightcurves for the target star (left) and field star (right), showing a deep event in the field star and only a very shallow event creeping into the target star pixels which demonstrates that the source of the signal is not the target star but the field star.
\label{fig:COSp_FP_LK}}
\end{figure*}

An example of a spurious centroid offset for a planet candidate (EPIC 210957318.01) is shown in the upper panels of Figure \ref{fig:COSp_}, where the out-of-transit photocenter is locked on the bright field star in the upper left corner of the aperture instead of on the target star and the PRF fit to the difference image returns the position of the target star marking it as the source of the transit signal. This target is listed as a confirmed planet on the NASA exoplanet archive NExScI (Mayo et al. 2018), and as a planet candidate with COSp (``Centroid Offset Spurious'') in our catalog. An example of a spurious centroid offset for a false positive (EPIC 211808055.01) is shown in the lower panels of Figure \ref{fig:COSp_}. Here, the photocenters of the out-of-transit and the difference images are both locked on the bright field star in the upper left corner of the target's aperture, and DAVE flags the target as a potential false positive based on a centroid offset. However, deeper investigation using the custom apertures and extracted lightcurves shown on Figure \ref{fig:COSp_PC_LK} and \ref{fig:COSp_FP_LK} demonstrate that the transit signal is coming from the field star and thus the target is listed in our catalog as a false positive with COSp.

Another example of a COSp is when the photocenter can be measured for only 2-3 points (typically for long-period candidates with few transits). As this is insufficient for a meaningful centroid analysis, such targets are dispositioned as planet candidates, with an added COSp comment. In cases where there are 3-4 centroid measurements (and thus a defined confidence interval), the centroid offset is at the ${\rm > 3\sigma}$ level of significance and on the order of a pixel or larger, we add an additional comment indicating that the target may be a false positive due to a potential CO. 


\subsection{Vetting Metrics: Flux Analysis}

By design, the DAVE pipeline is fast\footnote{Processing one K2 lightcurve for one target for one dataset takes 29 sec on a standard laptop computer.} and fully automated---it requires only a target list---easy to test and impartial. For flux-based vetting of K2 candidates, we analyzed several aspects of the time-series lightcurves to ensure that 1) the signal is plausibly astrophysical, i.e.~due to a planetary transit or stellar eclipse instead of a false alarm due to e.g.~instrument systematics, star spots, stellar pulsation; and 2) the target is not an eclipsing binary. We examine the same plots and information for every detrending available (i.e.~AGP, EVEREST, PDC and SFF).

\subsubsection{Modelshift Analysis}

\begin{figure}
\centering
\plotone{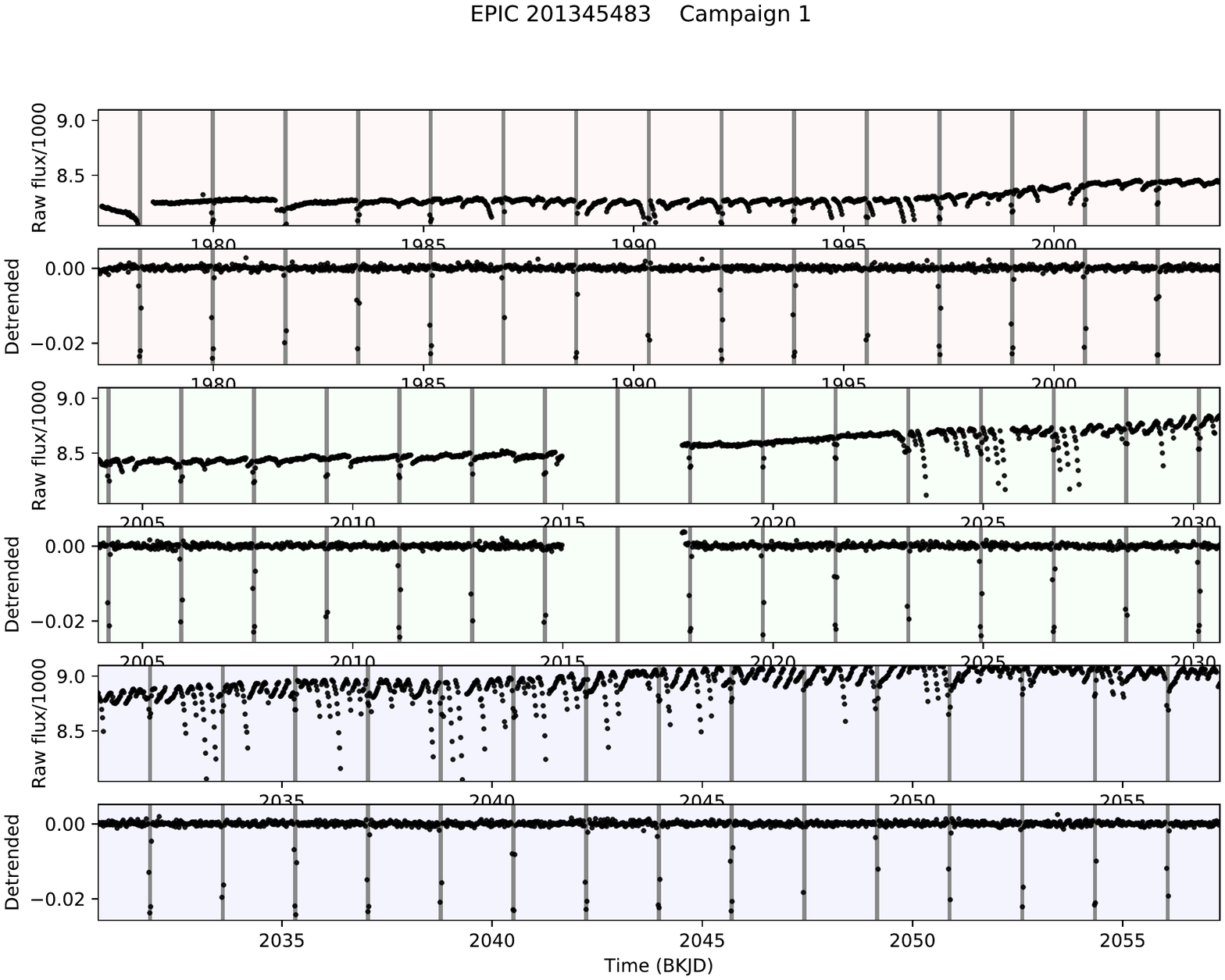}
\caption{Example for flux-based inspection of the entire EVEREST lightcurve of the planetary candidate EPIC 201345483.01. The first, third and fifth panels from top show the raw, SAPFLUX data. The second, fourth and sixth panels show the corresponding EVEREST data. The grey vertical lines indicate the transits of the candidate. There is no obvious reason for concern for any of the transits.
\label{fig:full_LC_201345483}}
\end{figure}

\begin{figure}
\centering
\plotone{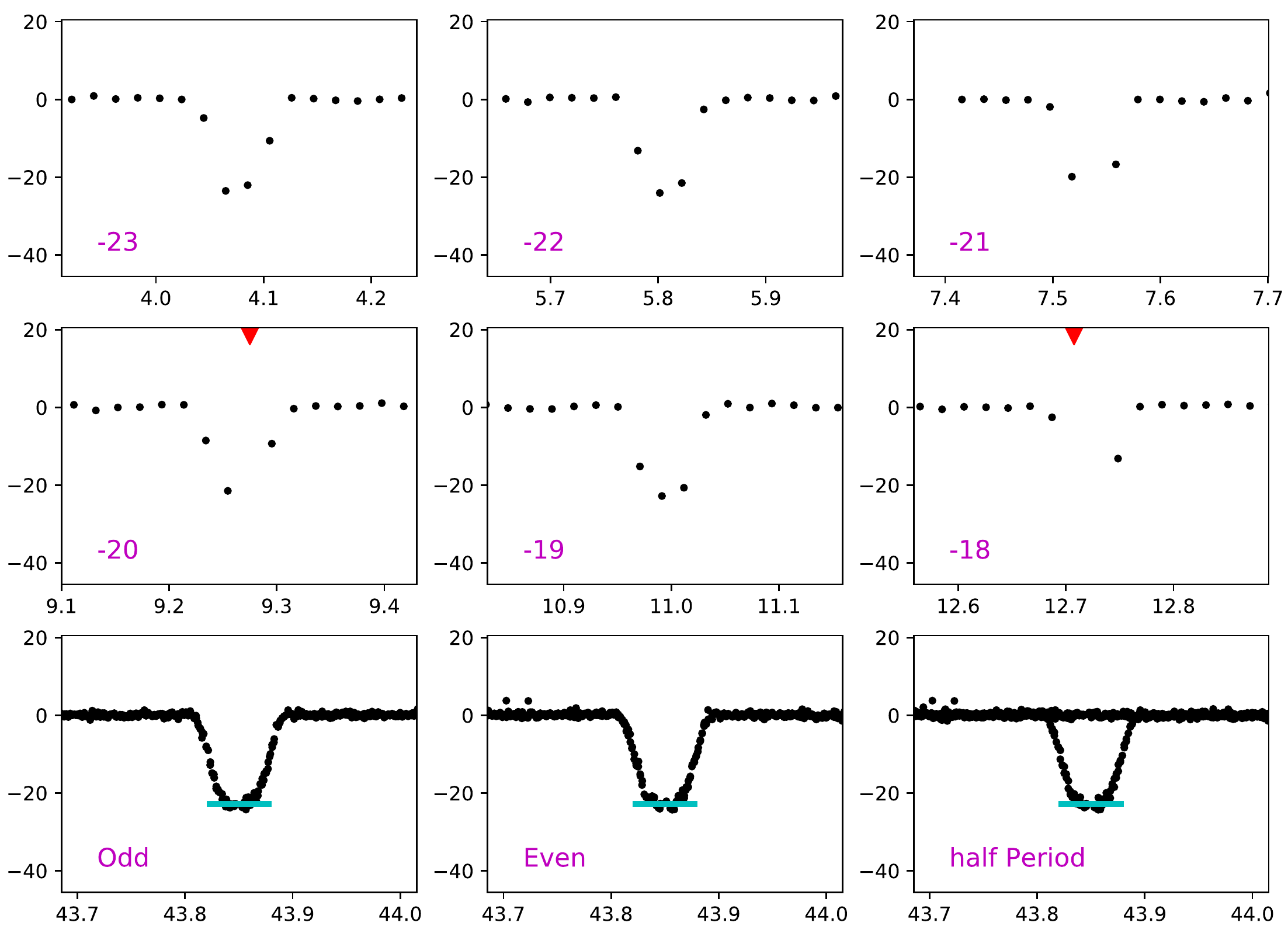}
\caption{Same as Figure \ref{fig:full_LC_201345483} but for individual transits (upper two rows) of EPIC 201345483.01. The bottom row shows the phase-folded odd (left), even (middle) and half-period  lightcurve. There is no significant difference between odd and even transits, and the listed period is correct.
\label{fig:indiv_201345483}}
\end{figure}

The flux-based analysis proceeds as follows. First, in order to minimize the effects of stellar variability, we median-filter the lightcurves with a window size of 50 points. Next, we inspect the phase-folded lightcurve and a zoomed-in plot of the primary transit.  We check whether the signal has a significant SNR compared to any out-of-transit variation and appears transit-shaped rather than V-shaped. Next, we check whether there is a secondary eclipse, or out-of-eclipse quasi-sinusoidal variation, which would indicate the object is an eclipsing binary. We then examine the entire photometric time-series---detrended and un-phased---with each identified transit highlighted according to the given ephemeris and duration. This is to ensure that the individual transit events do not occur near gap edges, during re-pointing events or other anomalies that may indicate the signal is systematic in origin instead of a bona-fide astrophysical signal. An example is shown in Figure \ref{fig:full_LC_201345483} for the EVEREST lightcurve of the planet candidate 201345483.01. Next, we inspect zoomed-in plots of all individual transits to check whether: 1) the shape and depth of each event is inconsistent with a transiting planet; 2) any transit exhibits an asymmetric depth profile; 3) a minority of the individual transits have significantly larger depths than the rest; 1), 2) and 3) would indicate that the feature is likely not a transit but instead has a systematic origin such as a sudden pixel sensitivity drop (SPSD). Next, we study the phase-folded lightcurve, focused on the primary event, separately for the odd- and even-numbered transits. A significant difference between the two indicates that the target is an eclipsing binary with an orbital period twice that of the detected period. Similarly, we also examine the phase-folded lightcurve focused on the primary at half of the detected period. If the signal appears coherent in this plot, it indicates that the signal is likely due to a transiting planet that was detected at twice the true orbital period. An example is shown in Figure \ref{fig:indiv_201345483} for the EVEREST lightcurve of the planet candidate 201345483.01.

\begin{figure}
\centering
\plotone{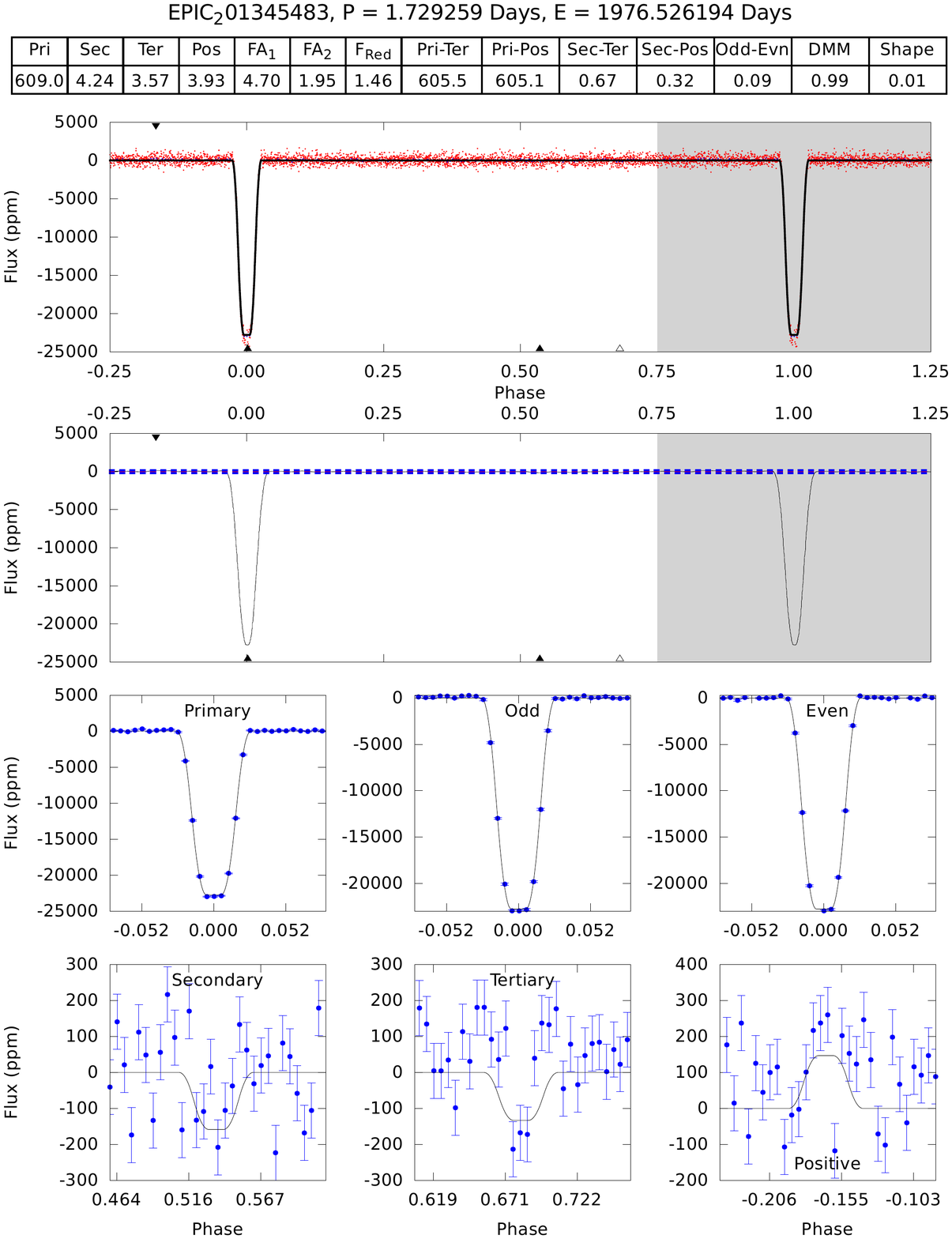}
\caption{Modshift results for planetary candidate EPIC 201345483. Upper two rows: folded and folded+convolved EVEREST lightcurve; Lower two rows: the individual panels, clockwise from upper left show the model fit to: all transits (label ``primary''); all odd transits (label ``odd'') ; all even transits (label ``even''); most prominent positive feature (label ``positive''); most prominent tertiary feature (label ``tertiary''); and most prominent secondary feature (label ``secondary''). There is no significant odd-even difference, no secondary or tertiary eclipses/transits, no positive features, and no sinusoidal modulations, and so this is a solid planet candidate. 
\label{fig:modshift_201345483}}
\end{figure}

\begin{figure}
\centering
\plotone{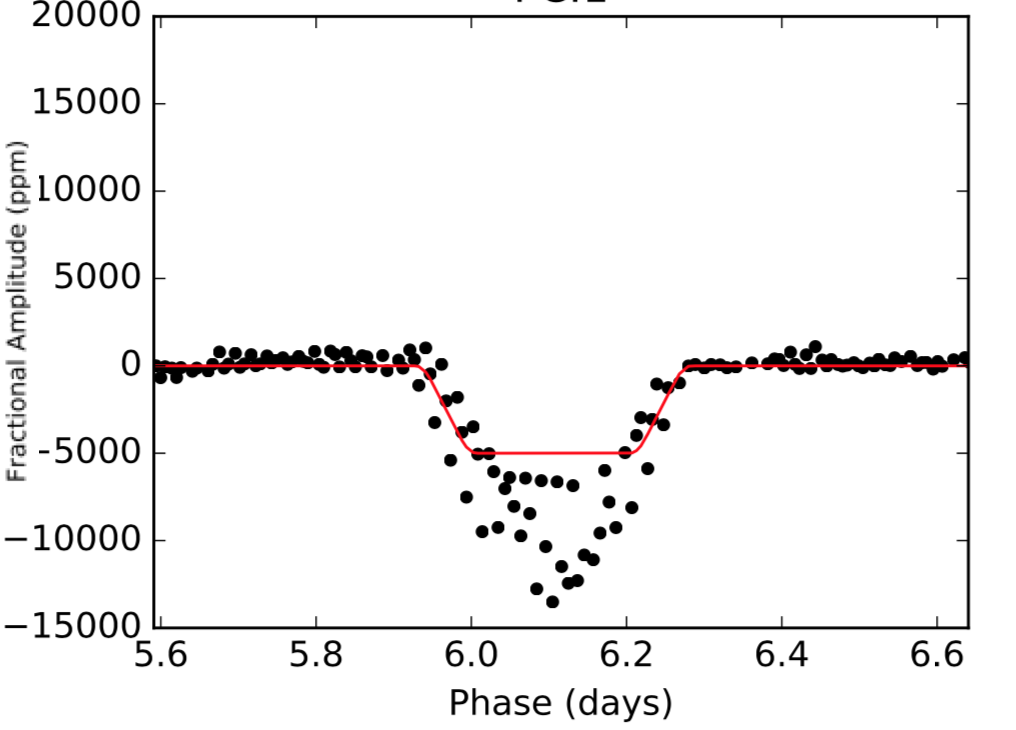}
\plotone{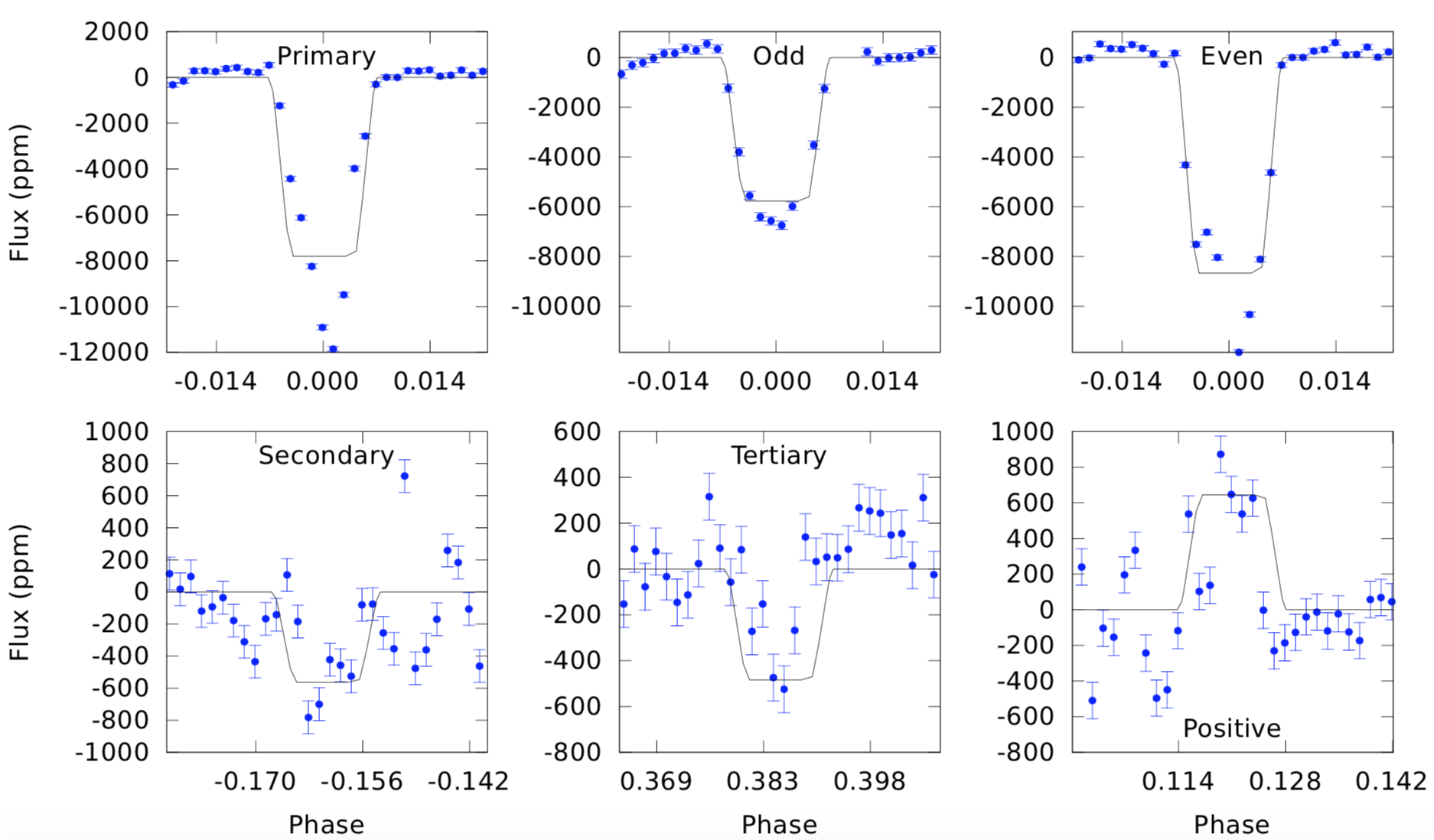}
\caption{Upper panels: Phase-folded lightcurve (AGP detrending). Lower panel: Modshift results for false positive EPIC 212443457.01. The target is a false positive due to a significant odd-even difference, indicating an eclipsing binary. We note that while Petigura et al. (2017) mark this target as a planet candidate they also comment that the transit is deep, irregular and the target is a ``possible hierarchical triple''.
\label{fig:modshift_212443457}}
\end{figure}

\begin{figure}
\centering
\plotone{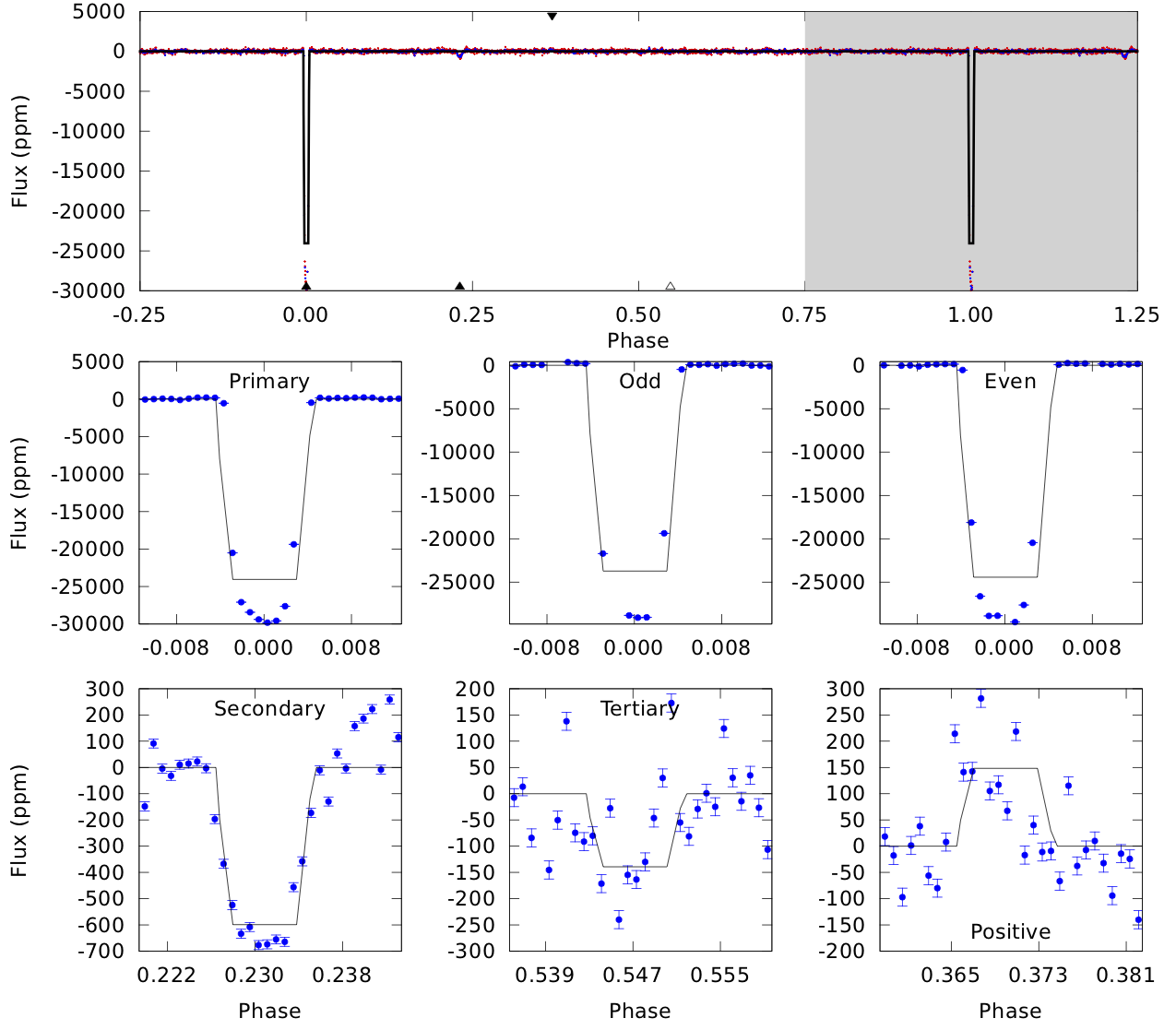}
\caption{Upper panels: Modshift results for false positive EPIC 214611894.01. Lower panel: Phase-folded lightcurve (AGP detrending). The target is a false positive due to a significant secondary eclipse, indicating an eclipsing binary.
\label{fig:modshift_214611894}}
\end{figure}

After careful inspection of each of these plots, we then examine the Modelshift plots (see A.3.4 of Thompson et al. 2018 for detailed description). Briefly, Modelshift is an automated procedure that convolves the transit model fit with the phase-folded lightcurve to highlight features that have similar shape, depth and duration to the primary transit event at phases other than 0.0. This allows a number of quantitative measurements. For example, the strength of the primary event is compared to the systematic red noise out of transit to ensure it is a significant detection. Additionally, the second strongest decrease in flux (aside from the primary) is compared to the systematic red noise, to test whether it can be a plausible secondary eclipse due to an eclipsing binary. Any positive events that might suggest that the system is a heartbeat star, self-lensing binary, or other non-planetary system are flagged and examined as well. Modelshift also performs a quantitative measurement of odd-even differences to examine eclipsing binaries detected at half their true orbital period, as well as a check on the consistency of the individual transit depths. Plots are provided to assist the user to determine whether the data qualitatively agrees with the quantitative measurements given. Example results from the Modelshift flux vetting are shown on Figures \ref{fig:modshift_201345483}, \ref{fig:modshift_212443457}, and \ref{fig:modshift_214611894}, demonstrating a planetary candidate (EPIC 201345483), a false positive due to odd-even difference (EPIC 212443457) and a false positive due to a significant secondary eclipse (EPIC 214611894). 

By examining the multiple facets of the photometric lightcurve described above, for every available detrending, and in a variety of qualitative and quantitative formats, we ensure that a target labeled as a candidate is plausibly due to a transiting planet. These metrics allow us to rule out false positives due to significant, un-detrended systematics, or eclipsing binaries with either a significant secondary or out-of-eclipse variations. For completeness, we also investigate the calculated size of the transiting object using the stellar radii listed in the catalog of Stassun et al. (2018), cautioning that these could be systematically off. With this caveat in mind, if the calculated size of the transiting object is larger than ${\rm 2R_{Jup}}$---i.e.~about twice the size of an M9V (e.g.~Kaltenegger \& Traub 2009) and thus potentially a star instead of a planet---the target is marked as a planet candidate but with an added flag of ``pVDE'' (``potential very deep eclipse'').

\begin{figure}
\centering
\plotone{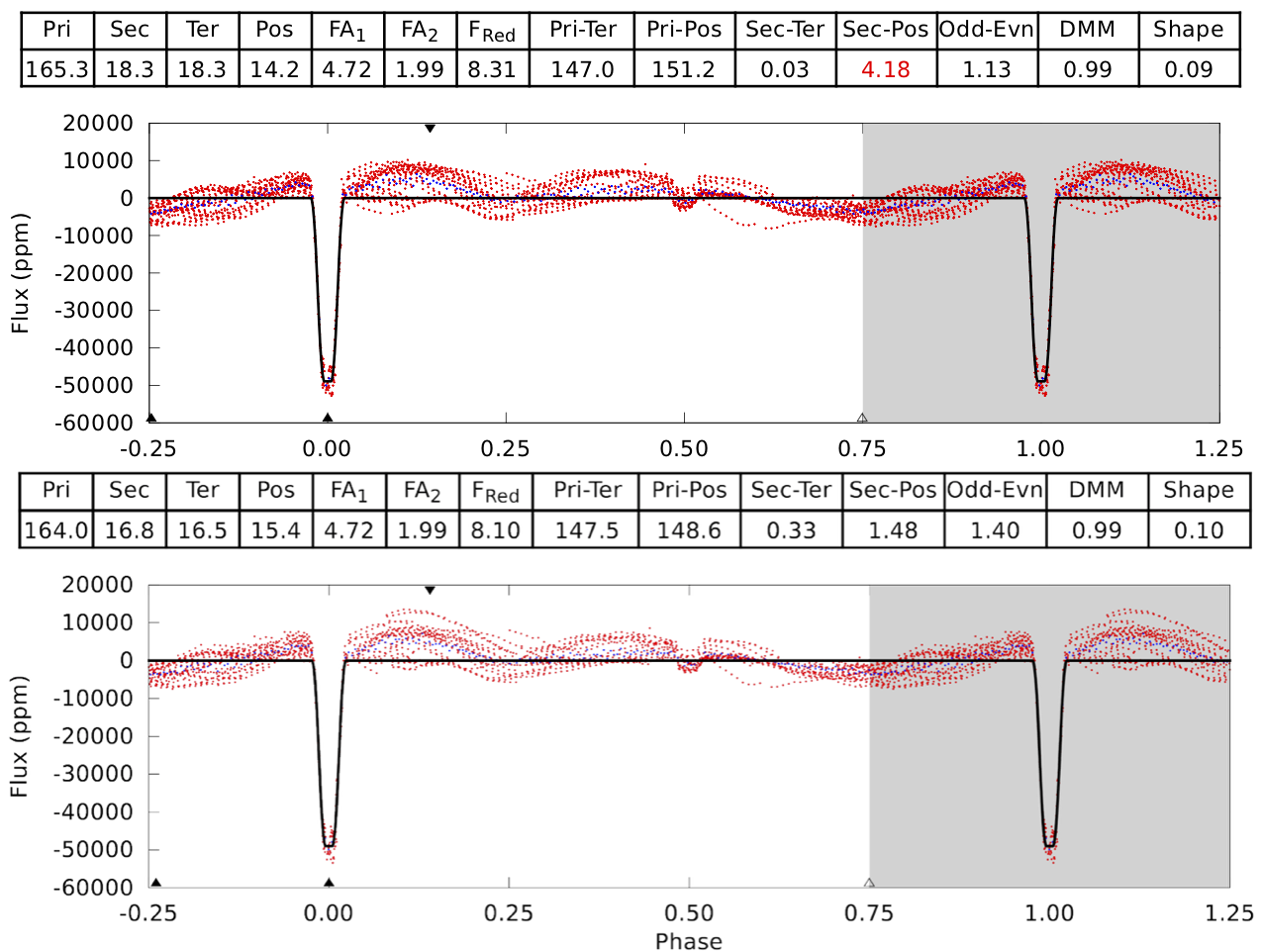}
\caption{An example of DAVE missing a clear secondary eclipse (EPIC 206135267.01), showing the results from the modshift analysis of the EVREST data (upper panel) and PDC data (lower panel). Systems like this demonstrate the benefit of complementary manual analysis. 
\label{fig:206135267_FP}}
\end{figure}

Additionally, to avoid potential problems related to particularly challenging candidates (e.g.~Shporer et al. 2017), to cases where the metrics are not well-tuned for K2 data, or to cases where the four dispositions for a particular target (based on the four different lightcurves) differ from each other, we complemented the Modelshift analysis by visual inspection of all DAVE dispositions. To minimize the introduction of human bias, the vetters look for the same features DAVE is analyzing, e.g.~odd-even difference, a secondary eclipse, transit-like shape, consistent transit depth and sufficient SNR. An example where the Modelshift module misses a clear secondary eclipse due to strong lightcurve variability yet human vetting captures it easily is EPIC 206135267, shown in Figure \ref{fig:206135267_FP}. The added benefit of human analysis is marking features that are not part of the vetting procedure but may nevertheless be astrophysically interesting (e.g.~flares, self-lensing events, etc.). Overall, if there is no sufficient reason to flag potentially doubtful targets as false positives, either based on DAVE's dispositions or on visual inspection, we aim to err on the side of passing them as planet candidates. 

Most of DAVE results are unambiguous, i.e. the different dispositions for a particular target are consistent with each other and visual inspection agrees with the automated vetting; these dispositions are referred to as ``accurate'' in Section \ref{sec:catalog}. Targets that required further scrutiny and discussion (85 in total) were inspected by at least two members of our team.  Of these targets, 51 were consistently identified as planet candidates by 3 vetters, and the rest as potential false positives. The dispositions where the human vetters disagree with DAVE are marked as ``not convincing'' and marked in the DAVE catalog as ``NC''. Overall, of the 2886 individual dispositions produced by DAVE, visual inspection agrees for 2609 cases, or about 90\%.

\subsubsection{Transit-like Analysis}

It is not uncommon for transit-searching algorithms to return short-period, quasi-sinusoidal false positives that are caused by stellar spots or contact eclipsing binaries. To detect these we use LPP (Locality Preserving Projections)---a transit-like metric that uses dimensionality reduction and K-nearest neighbors to distinguish transit-looking folded lightcurves from the rest. The LPP algorithm currently implemented by DAVE is the same algorithm as that used for the DR24 KOI Catalog vetting (See Section 3.2.1.1 of Coughlin et al. 2016 and Thompson et al. 2016), and we outline it here for completeness. We note that LPP is similar to the method used by Matijevic et al. (2012), where a local linear embedding is used to distinguish between detached and contact eclipsing binaries. LPP, however, differs as it can be applied to Threshold Crossing Events (TCEs) with parameters outside those of the training set. The output from the LPP transit-like metric (TLM) is a single number representing the degree of similarity between the shape of a TCE shape to the shape of known transits. The method works as follows. 

First, the lightcurve is folded and binned down to 141 data points. The binning is not uniform but instead it emphasizes the points around the transit. The algorithm then uses LPP dimensionality reduction (He \& Niyogi 2004) to reduce those 141 binned points down to 20. Next, using the k-nearest neighbors in this lower-dimensionality space, the distance to the 15 closest transit signals are averaged. This average distance in 20 dimensions is the value given for the TLM.  Lower values indicate that the signal in question is clustered near known transit-like signals and thus it is likely shaped like a transit.  DAVE's TLM relies on the mapping to lower dimensions generated from the population of known transits developed for the DR24 Catalog (Coughlin et al. 2016).  We note that signal detrending can have an impact on the effectiveness of the metric and so we feed the TLM median detrended lightcurves.

\begin{figure}
\centering
\plotone{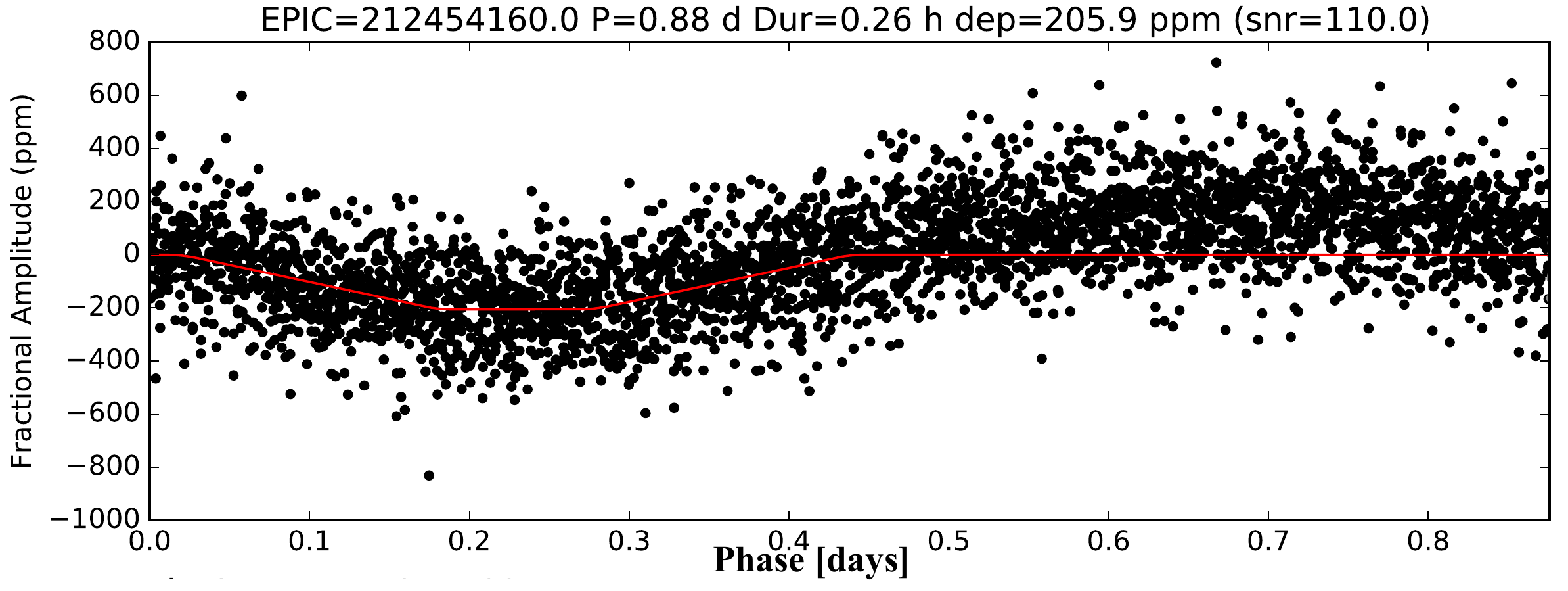}
\caption{An example of the TLM vetting module of DAVE, demonstrating quasi-sinusoidal modulations masquerading as a transit signal (EPIC 212454160.01). The data shown represents the phase-folded AGP lightcurve.
\label{fig:lpp_}}
\end{figure}

An example result from the TLM module is shown in Fig. \ref{fig:lpp_}, demonstrating a false positive where the signal is consistent with quasi-sinusoidal modulations instead of a transit (EPIC 212454160). 

\section{A Benchmark Catalog of Uniformly-Vetted K2 planet candidates}
\label{sec:catalog}

We used the DAVE pipeline to vet the K2 Planet Candidate Catalog hosted on the NASA Exoplanet Archive as of Aug 6, 2018, utilizing the four publicly-available lightcurves---AGP, EVEREST, PDC, and SFF---that are available as high-level science products at the Mikulski Archive for Space Telescopes (MAST). Capitalizing on this treasure trove of data, we produced a uniformly-vetted, publicly-available DAVE catalog of \alltargets planets using data from K2 Campaigns 1 through 10\footnote{AGP does not cover Campaign 1.}. 

The catalog lists the EPIC ID of each K2 planet candidate, the transit properties, disposition (planet candidate ``PC'' or false positive ``FP''), the reason for the disposition (e.g.~centroid offset during transit, as ``CO''), as well as additional comments (for example, presence of field stars in the target's aperture, as ``FSAp''). DAVE dispositions for planet candidates are uploaded to the Kepler Threshold Crossing Event Review Team website (http://keplertcert.seti.org/DAVE/) and to the Exoplanet Follow-up Observing Program (ExoFOP) website hosted by NExScI. Supporting materials, including the searched lightcurves, will also be provided for each EPIC ID. Overall, the catalog is a community-facing product designed to serve as a multi-purpose tool for e.g.~A) providing quick-look results for interesting targets (if the lightcurves are available at MAST); B) minimizing or even completely removing the need to wait for an accepted publication since we will continue to update the DAVE website; C) facilitating exoplanet confirmation and characterization; and D) enabling the community to make their own judgment about a particular candidate. 

\begin{figure*}
\centering
\plotone{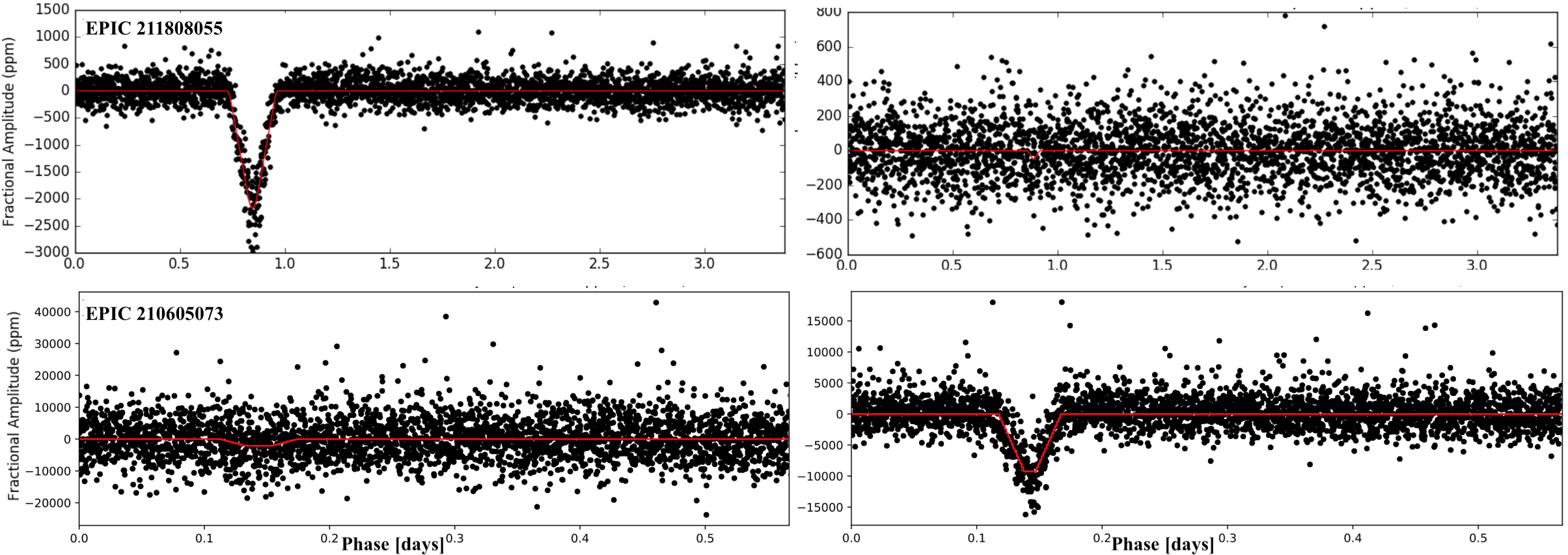}
\caption{Comparison between AGP (left panels) and EVEREST (right panels) lightcurves for EPIC 211808055.01 (upper panels) and EPIC 210605073.01 (lower panels) in terms of normalized flux as a function of orbital phase. DAVE marks EPIC 211808055.01 as a planet candidate in AGP data and a false positive in EVEREST data due to low SNR, and vice versa for EPIC 210605073.01. The disposition for the former target is accurate for AGP data and inaccurate for EVEREST data, and vice versa for the latter target. EPIC 211808055.01 is listed in our catalog as a false positive due to centroid offset (see Figures \ref{fig:COSp_} and \ref{fig:COSp_FP_LK}), and EPIC 210605073.01---as a planet candidate.
\label{fig:agp_vs_everest_}}
\end{figure*}

DAVE also provides a utility platform for comparison between lightcurves produced by different teams, each employing a different approach to systematics correction on a star-by-star basis. For example, when comparing the vetting results for AGP, EVEREST, SFF, and PDC our analysis shows that about one in three planet candidates do not have sufficient signal-to-noise ratio (${\sim5-10}$, depending on the target) in at least one lightcurve set. An example of a low SNR in one lightcurve but a clear transit in another is demonstrated in Figure \ref{fig:agp_vs_everest_} where we compare AGP and EVEREST data for the same target. For the case of EPIC 211808055.01, the transit signal is clearly defined in AGP data and has a low SNR in EVEREST data\footnote{Note that this target is a false positive because the transit signal is coming from a field star, see Figure. \ref{fig:COSp_FP_LK}}; for EPIC 210605073.01 it is the other way around. In general, the dispositions can vary between the different lightcurves on a target-by-target basis, and thus vetting multiple datasets for each target is crucial to distinguish between a bona-fide planet candidate and a false positive. 

\begin{figure*}
\centering
\plotone{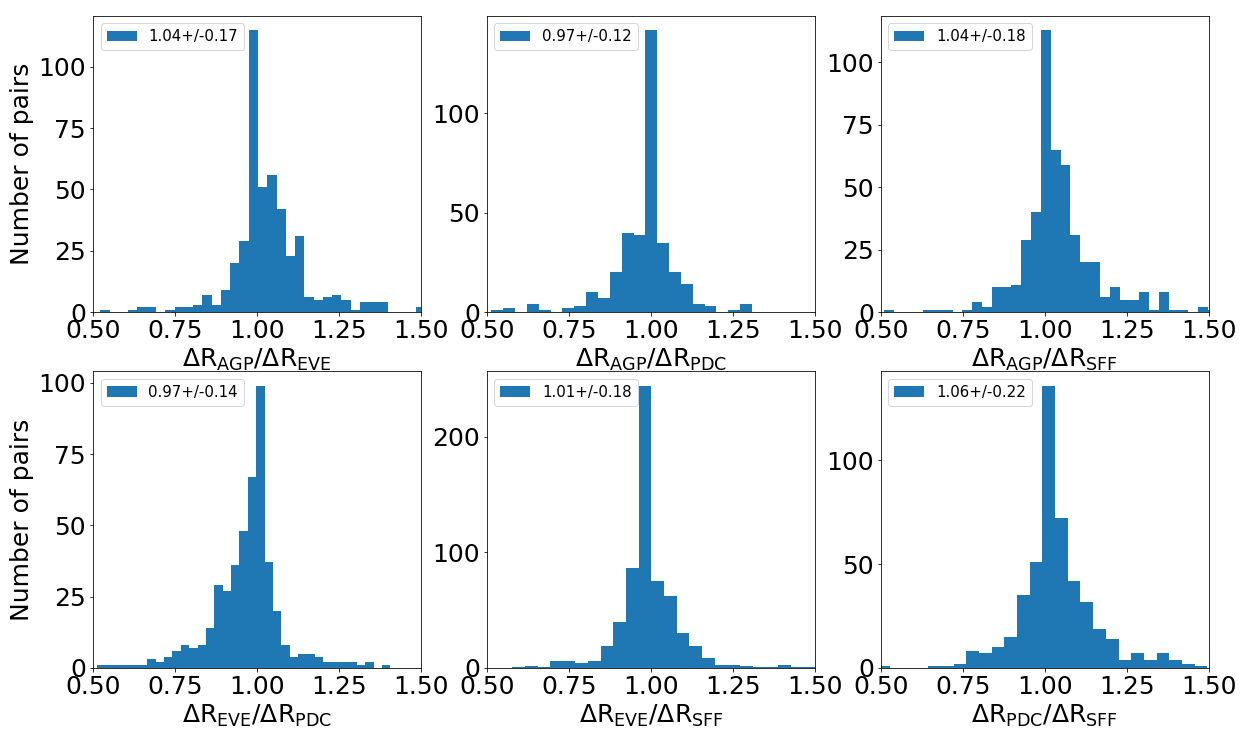}
\caption{Ratio of planet-star radius ratios, e.g. ${\rm (R_{planet}/R_{star})_{AGP}/(R_{planet}/R_{star})_{EVE}}$, for the candidates that show significant transits in the respective pair of datasets. The legends list the corresponding mean and ${\rm 1\sigma}$ values. We do not observe a trend with radius ratio and light curve detrending pipelines.
\label{fig:RadRatRat_}}
\end{figure*}

For candidates whose transits have sufficient signal-to-noise ratio in at least two datasets, we find that the calculated radius ratios (${\rm \Delta R = R_{planet}/R_{star}}$) are consistent across the four detrendings (AGP, EVEREST, PDC, and SFF). We compare the radius ratios between the different detrendings, in terms of ${\rm \delta R = (\Delta R_{AGP}) / (\Delta R_{EVE})}$, in Figure \ref{fig:RadRatRat_}. The number of pairs are different because a candidate can have significant transits in as few as two datasets or in as many as four---and on a target-by-target basis. We note that there are several outliers where targets have ${\rm \delta R >> 5}$ between two datasets yet the transits are significant in both, e.g. EPIC 211831378 where the transit depth in PDC is ${\sim0.1}$ and in SFF it is ${\rm \sim300 ppm}$. There is no obvious reason for such differences.

\begin{figure*}
\centering
\plotone{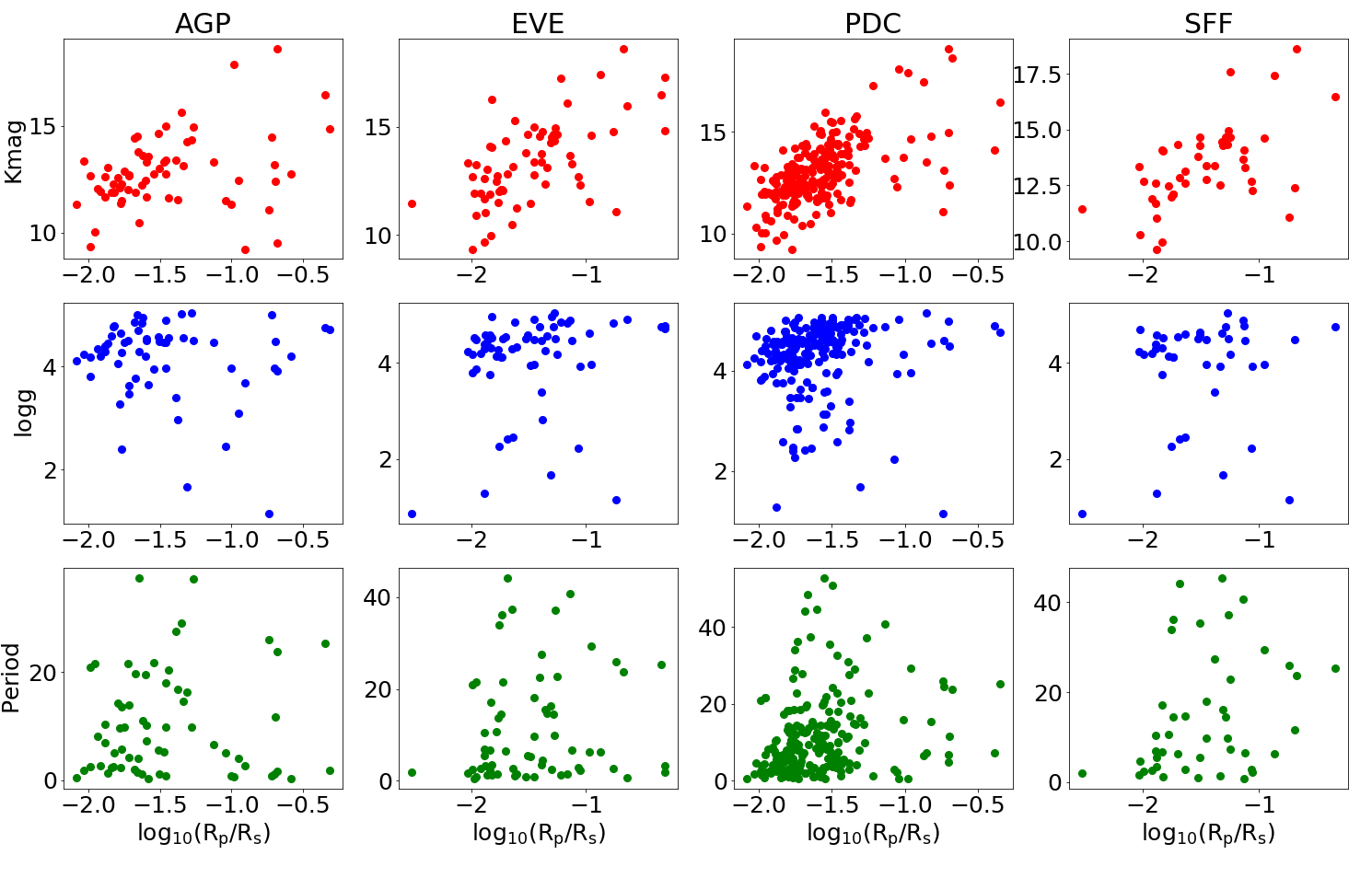}
\caption{Kepler magnitudes, stellar gravities, and orbital periods as a function of ${\rm R_{planet}/R_{star}}$ (as listed on NExScI) for the candidates that do not show significant transits in a particular detrending pipeline.
\label{fig:no_transits_}}
\end{figure*}

The number of candidates showing transits with sufficient SNR in each detrending pipeline is listed in Table \ref{tab:RMN_}. Our analysis shows that ${\sim90\%}$ of the candidates show significant transits in AGP, EVE and SFF data, and ${\sim70\%}$ in PDC data. Overall, ${\sim 67\%}$ of the targets show significant transits in all four lightcurve sets, ${\sim 23\%}$ in three lightcurve sets, ${\sim 5\%}$ in two lightcurve sets, ${\sim 3\%}$ in one lightcurve set, and ${\sim 2\%}$ do not show a significant transit in any set. The orbital periods, Kepler magnitudes, and stellar gravities of the candidates showing no significant transits in a particular detrending pipeline are shown in Figure \ref{fig:no_transits_} as a function of the candidate's radius ratio as listed on NExScI. Overall, we find that there is no one-size-fits-all recipe for analyzing these candidates. Thus users are recommended to examine as many different detrendings as possible when evaluating individual systems and modeling them. 
 

\subsection{Comparison between NExScI and DAVE dispositions}

Of the \alltargets K2 planet candidates we investigated, NExScI listed \NExScIConfirmed as confirmed planets, \NExScIFalse as false positives and \NExScICandidates as planet candidates (as of Aug, 1, 2018). We note that 8 targets---EPIC 201324549.01, 205990339.01, 206432863.01, 210414957.01, 210754505.01, 211804579.01, 212572452.01, and 228729473.01---are listed as both a planet candidate and as a false positive at the time of writing (e.g. Livingston 2018, and references therein). In addition, EPIC 206135267.01 is listed on NExScI as a planet candidate with a reference to Crossfield et al. (2016) and Vanderburg et al. (2016) yet it is marked as an ``Obvious Binary'' by Crossfield et al. (2016) (their Table 3). Another target, EPIC 212443457.01 is also listed as a planet candidate on NExScI but is marked as a ``irregular transit shape'' by Petigura et al. (2017).  

Our analysis of this sample identifies \PCs as planet candidates and \FPs as potential false positives. These are distributed as described below.

\subsubsection{DAVE Analysis of Confirmed Planets}

All \NExScIConfirmed targets listed on NExScI as confirmed planets pass our vetting analysis and are marked in the DAVE catalog as planet candidates (``PC''). Their corresponding DAVE dispositions are listed in Table \ref{tab:CP}. We note that we vet targets independent of any prior knowledge of whether the system is confirmed or not.

Four of the confirmed planets show signs of potential---but not prominent enough to declare false positive---photocenter shift and although they are dispositioned in our catalog as planet candidates we recommend further investigation. In addition, while DAVE's automated vetting passes EPIC 212554013.01, visual inspection of the ModShift results indicates that there may be a potential secondary eclipse, very weak (depth of ${\rm \sim100-200}$ parts per million), identified by the module at the same phase (${\rm \approx0.38}$) for the AGP, EVEREST and SFF lightcurves, albeit not as a statistically significant feature. However, as the (primary) transit depth indicates a Jupiter-sized object (${\rm \sim10R_\oplus}$) and thus the potential secondary feature may be an occultation, and the depth of the potential secondary feature is comparable to the height of positive features present in the lightcurve, we list this target as a planet candidate. 

\begin{figure*}
\centering
\plotone{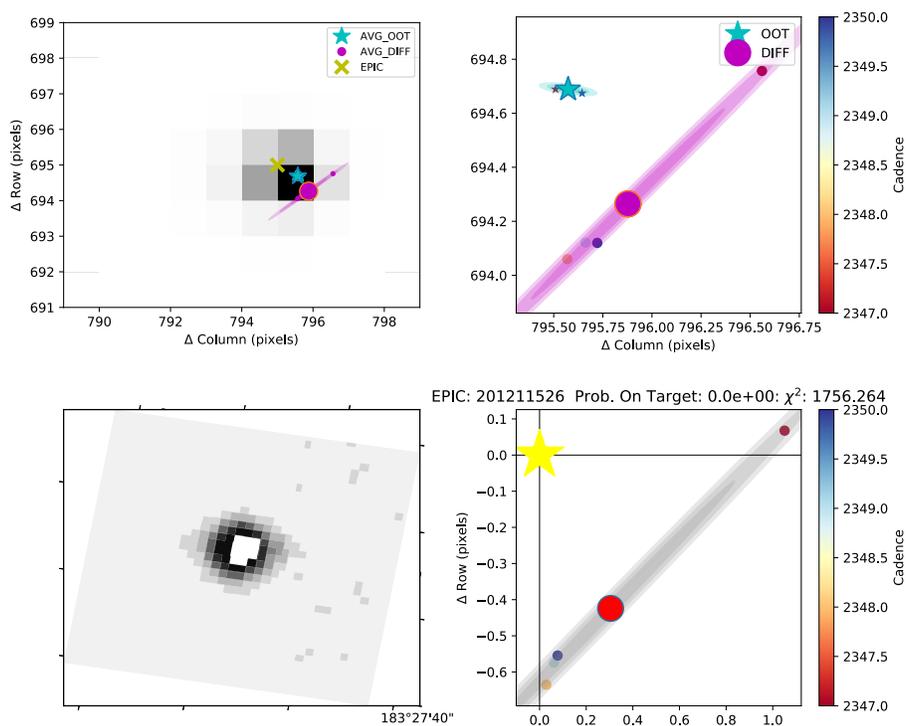}
\caption{Measured out-of-transit and difference image centroid positions for EPIC 201211526 (results for PDC data), listed as a confirmed planet on NExScI but indicating a potential centroid offset in DAVE of ${\sim0.5}$ pixel with a ${\chi^2\approx1800}$. Lower left panel shows a 1\arcmin x 1\arcmin~2MASS J-band image. 
\label{fig:201211526_pCO}}
\end{figure*}

\begin{figure*}
\centering
\plotone{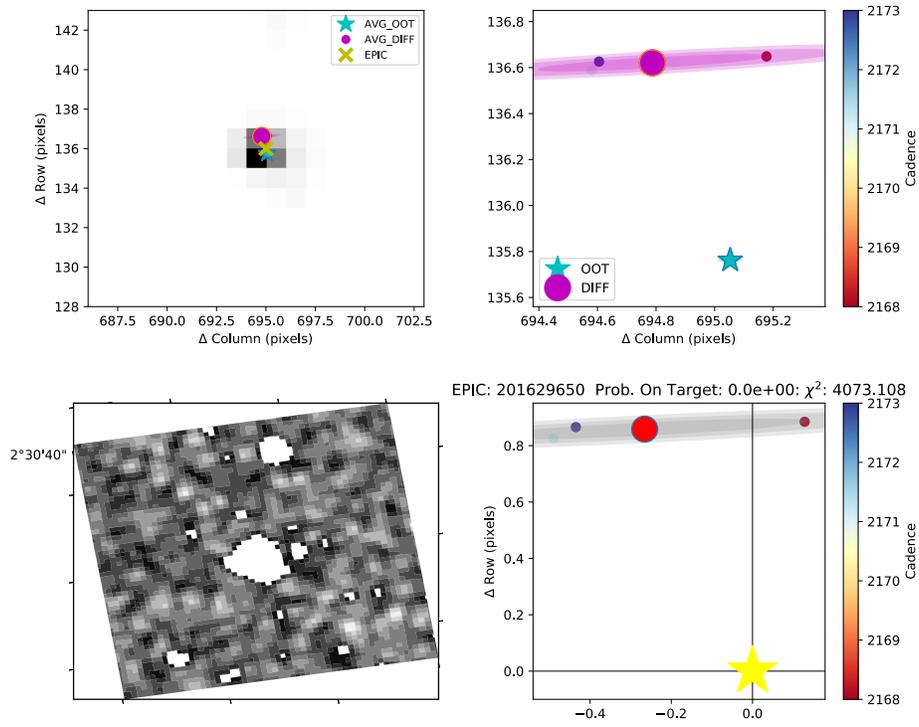}
\caption{Same as Figure \ref{fig:201629650_pCO} but for EPIC 201629650. The measured offset is ${\sim1}$ pixel with a ${\chi^2\approx4000}$. Vetting results for EVEREST data.
\label{fig:201629650_pCO}}
\end{figure*}

\begin{figure*}
\centering
\plottwo{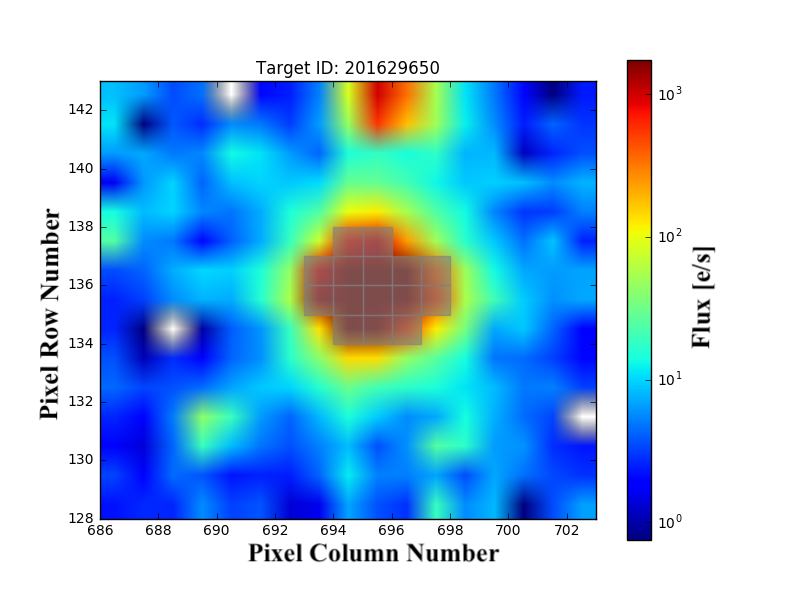}{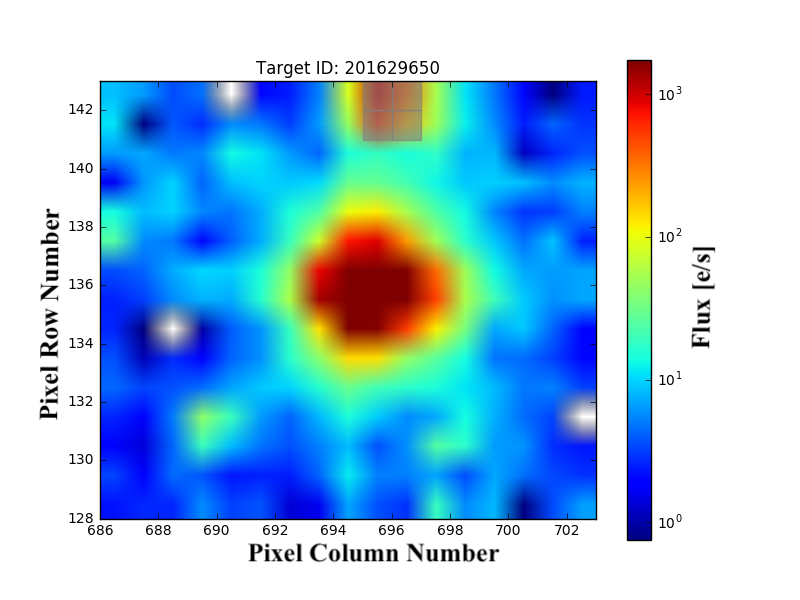}
\plottwo{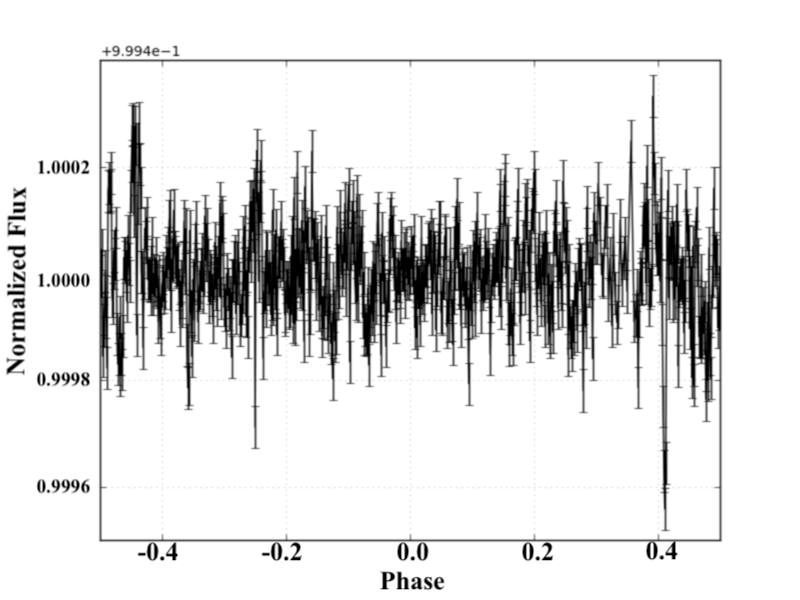}{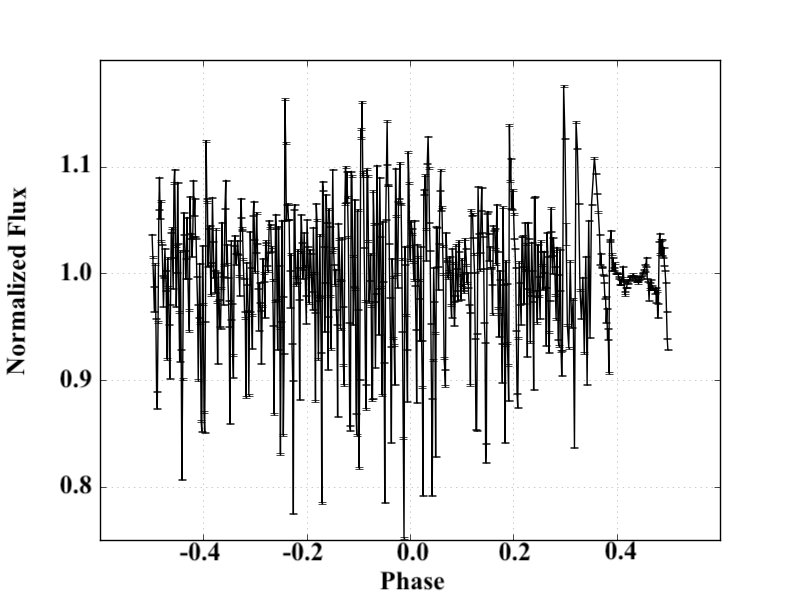}
\caption{Custom apertures (upper panels) and corresponding lightcurves (lower panels) for EPIC 201629650 using lightkurve. While the target's aperture (upper left) does produce the transit feature (lower left panel, near phase 0.4), the systematics are poorly removed from the lightcurve corresponding to the off-target aperture (right panels) and we cannot rule it out as a potential source of the observed transits.
\label{fig:201629650_LK}}
\end{figure*}

\begin{figure*}
\centering
\plotone{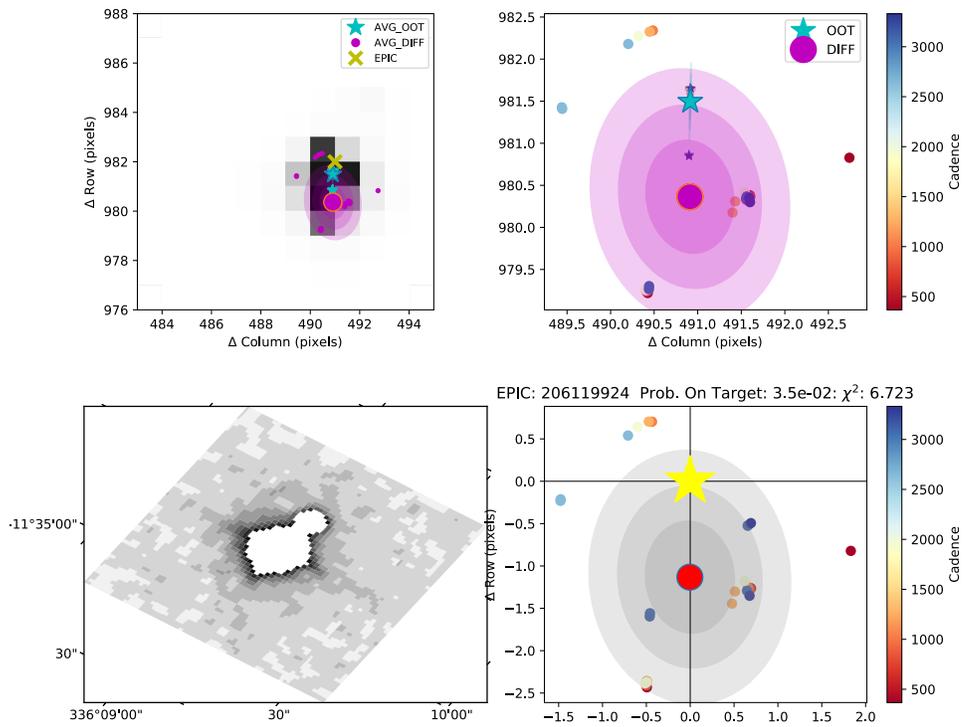}
\caption{Same as Figure \ref{fig:201629650_pCO} but for EPIC 206119924, listed as a confirmed planet on NExScI yet showing a potential centroid offset in DAVE of ${\approx1}$ pixel at a ${\rm \sim2\sigma}$ level of significance with a ${\chi^2\approx7}$. Vetting results for AGP data.
\label{fig:206119924_pCO}}
\end{figure*}

\begin{figure*}
\centering
\plotone{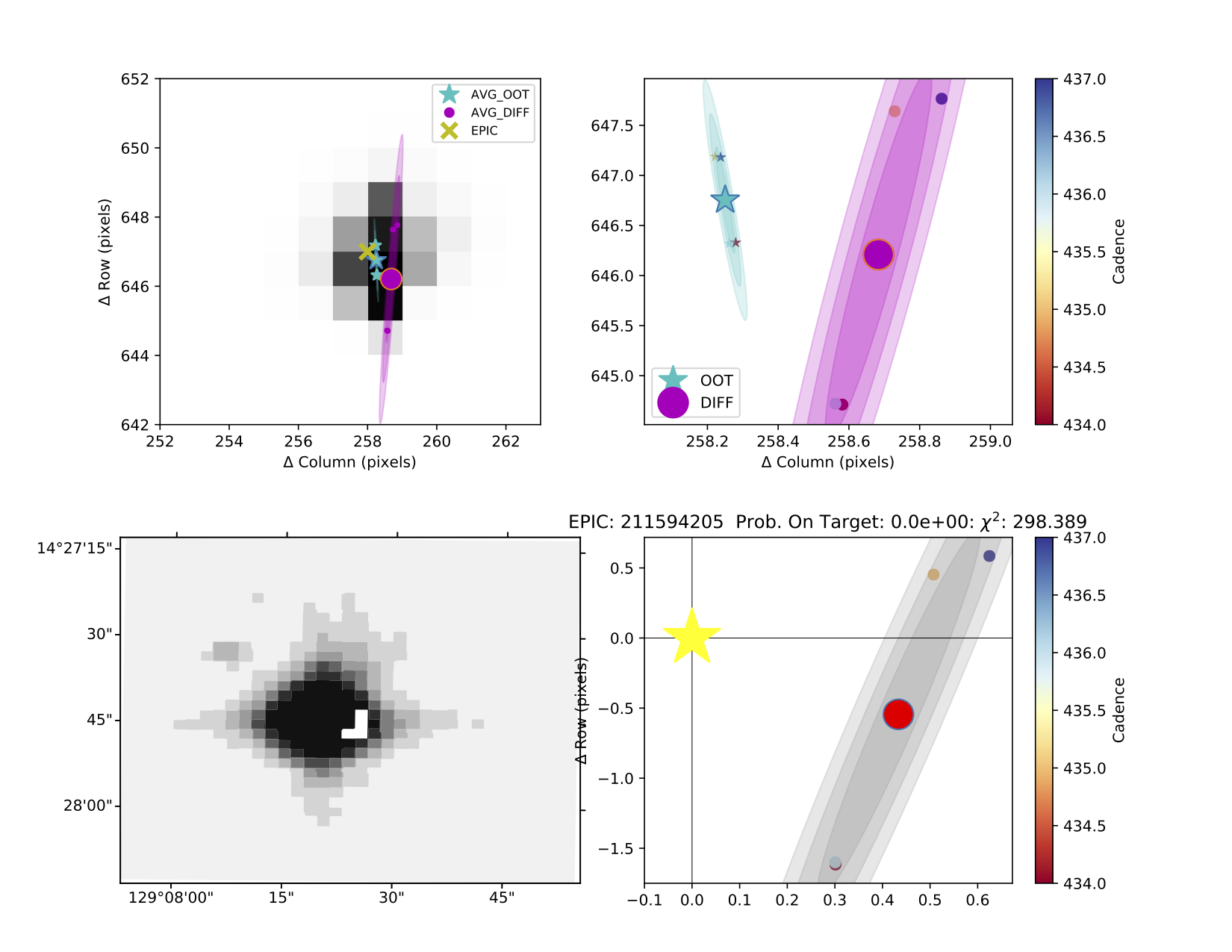}
\caption{Same as Figure \ref{fig:211594205_pCO} but for EPIC 211594205.01, listed as a confirmed planet on NExScI yet showing a potential centroid offset in DAVE of ${\sim0.7}$ pixels with a ${\chi^2\approx300}$. Lower left panel shows a 1\arcmin x 1\arcmin~DSS Red image (the faint companion above and to the left of the target is not visible on 2MASS images). Vetting results for AGP data.
\label{fig:211594205_pCO}}
\end{figure*}

The four targets with potential centroid offsets are as follows: \\


i) EPIC 201211526.01: See Figure \ref{fig:201211526_pCO}. The centroid offset is most pronounced in the vetting analysis of the PDC data---showing 4 cadences with measured centroids---and less so in the other three lightcurve sets where there are 3 measured centroids. Query of the 2MASS catalog does not reveal any obvious field stars within the K2 aperture (lower left panel of the figure). \\

ii) EPIC 201629650.01: See Figure \ref{fig:201629650_pCO}. There are two field stars North and South of the target, the former much brighter. DAVE measures a ${>3\sigma}$ centroid offset in a northern direction, although it is based on 3 centroid positions only. Deeper investigation of this target using custom apertures with the lightkurve package (Vinicius et al. 2018) was inconclusive (see Figure \ref{fig:201629650_LK}). Nevertheless, given that the magnitude of the measured centroid offset is nearly a pixel and the direction is towards the brighter of the field stars, we add a comment indicating a potential centroid offset. \\


iii) EPIC 206119924.01: See Figure \ref{fig:206119924_pCO}. The measured centroid offset is at the ${\sim2\sigma}$ level, and the individual difference image photocenters are scattered across the target's aperture. Closer inspection of the 2MASS J-band image (lower left panel) indicates there is a faint nearby field star North of the target. Two stars are blended within the central 2-3 pixels and we could not employ custom lightcurve extraction to study the photometry of each star individually. The measured difference image photocenter is in the opposite direction of the field star which suggests that the offset may be spurious. However, for consistency within our catalog and despite its previous confirmation, we list the target as RFS for ``recommend further scrutiny''. \\

iv) EPIC 211594205.01: See Figure \ref{fig:211594205_pCO}. The measured centroid offset is ${\sim0.7}$ pixels at the ${>3\sigma}$ level, although it is based on four points. While there are no obvious field stars in 2MASS images, there is a faint field star SE of the target on DSS images, in the opposite direction of the measured photocenter shift. The two stars are blended within the central 2-3 pixels and we could not employ custom lightcurve extraction to study the photometry of each star individually. Thus we recommend further scrutiny of this target.

\subsubsection{Potential False Positives}

Of the \FPs targets flagged as potential false positives in the DAVE catalog (see Table \ref{tab:FP_final}), \NewFPs are listed in NExScI as planet candidates and \OldFPs as false positives. Our dispositions and the corresponding number of targets are listed in Table \ref{tab:FP_final}. The most common reason for flagging a target as a false positive (\FPsSigSec targets) is the presence of a secondary eclipse (indicating the target is probably an EB). Next is a measured photocenter shift during transit (\FPsCO targets, indicating the target star is not the source of the signal). \FPsOOTMOD targets exhibit out-of-transit modulations in phase with the detected transits (suggesting an EB); \FPsLCMOD targets exhibit non-transit like features (i.e.~stellar variability mimicking as transits), and \FPsOE targets show differences in the measured depths of odd and even transits (indicating an EB at half the proposed period). \FPsMoreThanOne targets exhibit more than one false positive indicator. 

We note that while DAVE detected secondary eclipses for two targets, EPIC 206036749.01 and EPIC 211705654.01, the depths of the corresponding primary transits indicate a Jovian or smaller orbiting body and the orbital periods are ${\sim2-4}$ days. As a result, these secondary eclipses may in fact be planetary occultations and the targets are thus listed in our catalog as planet candidates. Additionally, the calculated transit depth for \FPsVDE targets suggests a transiting object larger than ${\rm 2R_{Jup}}$, and \FPsVSHAPE targets show V-shaped transits---both potentially indicative of an EB. These targets are listed in our catalog as planet candidates with an added flag for respectively ``potential very deep eclipses'' or ``V-shaped transits''. 

\begin{figure*}
\centering
\plottwo{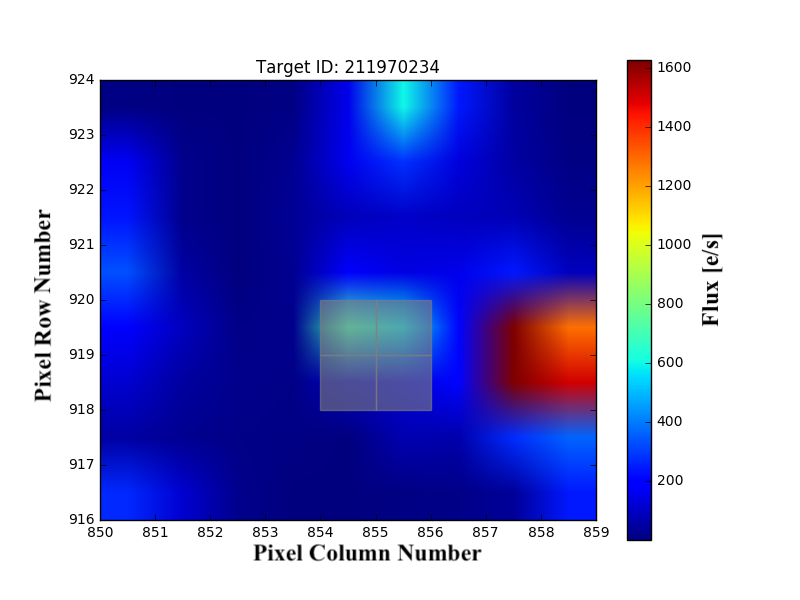}{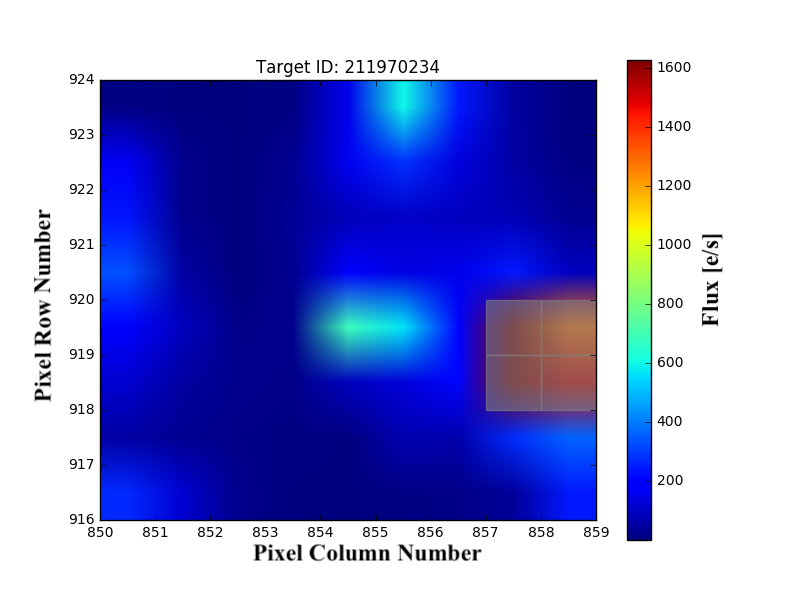}
\plottwo{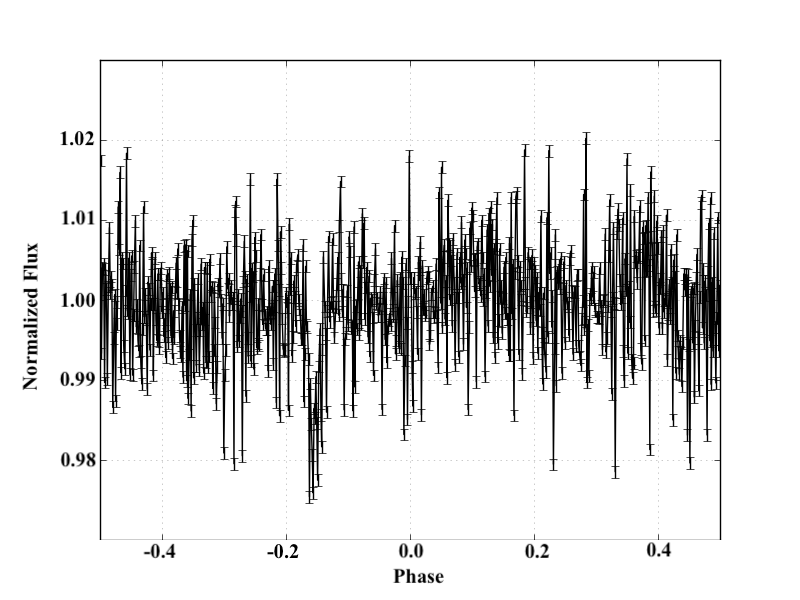}{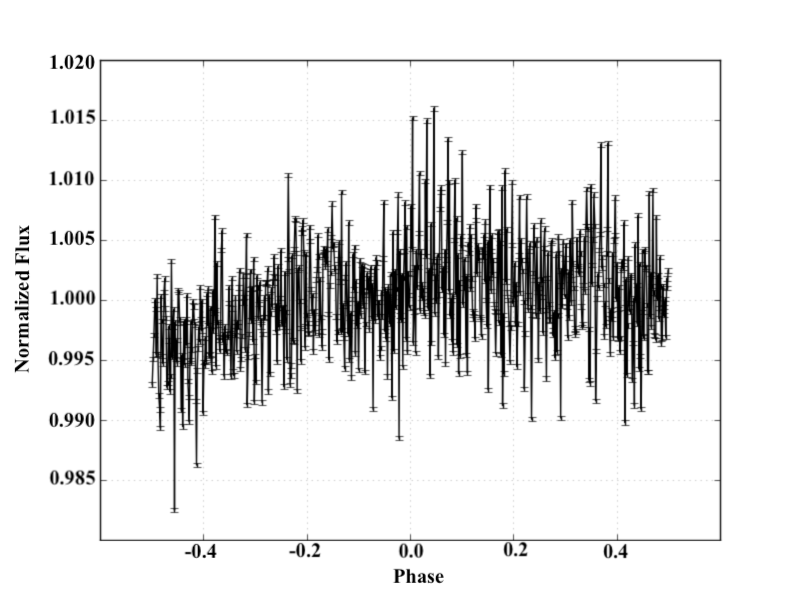}
\caption{Same as Figure \ref{fig:COSp_FP_LK} but for EPIC 211970234.01, listed as a false planet on NExScI due to ``inconsistent transit depth or blended photometry'' (Dressing et al. 2017). Using lightkurve, we confirmed that the transit signal is coming from the target star (left panels) and not from the nearby bright field star (right panels), indicating that the target is a bona-fide planet candidate. 
\label{fig:211970234}}
\end{figure*}

We note that while \OldFPsnotFound of the remaining NExScI false positives pass the automated vetting and are marked as planet candidates in our catalog, \OldFPsnotFoundbutFlagged are flagged as ``potential very deep eclipses'' or ``V-shaped transits'', and visual inspection marks two of them as likely false positives due to potential odd-even differences and/or out-of-transit modulations, in agreement with the results of Adams et al. (2016)\footnote{These two were not specially selected for further scrutiny---all DAVE results are inspected by at least one person, regardless of their automatic disposition.}. \OldFPsnotFoundnotFlagged are discussed in more details below.

i) EPIC 201324549.01 is listed as a false positive and flagged as a ``triple star system'' in Crossfield et al. (2016) but as a planet candidate by Barros et al. (2016) and Vanderburg et al. (2016). This target passes all our vetting tests and is marked in the DAVE catalog as a planet candidate. 

ii) EPIC 203581469.01 has a ${\rm SNR < 7}$ and is thus automatically flagged as a false positive by DAVE. As described below, for such targets we use a default disposition of planet candidate. 

iii) EPIC 205990339.01 is also listed as a planet candidate by Vanderburg et al. (2016) but a false positive with a probability of 1 by Crossfield et al. (2016); DAVE marks this target as a false positive in SFF and EVEREST data due to odd-even differences. However, visual inspection marks these as spurious since the period is long, there are only three transits, the transit has a short duration, and thus not well-sampled. 

iv) EPIC 206065006.01 has a ${\rm SNR < 4}$. By default we mark such targets as planet candidates even though DAVE flags them as false positives.

v) The two planet-system EPIC 206432863 (in a 2:1 resonance) is listed in NExScI as confirmed by Crossfield et al. (2016) but refuted by Shporer et al. (2017) based on RV measurements. DAVE flags the inner planet candidate as a false positive due to odd-even differences and centroid offset but visual inspection does not find these convincing given the intrinsic lightcurve modulations and the scatter in the measured photocenters, and flags the outer planet candidate as a false positive due to secondary eclipses---again disproved by visual inspection since these are the transits of the inner candidate.  

vi) EPIC 210414957.01 and EPIC 210754505.01 are listed on NExScI as both false positives due to out-of-transit modulations (Adams et al. 2016), and planet candidates (Barros et al. 2016, Crossfield et al. 2016). Visual inspection of these two suggests that there may indeed be potential out-of-transit modulations in AGP and EVEREST data, as well as a potential odd-even difference for EPIC 210754505.01 in EVEREST data (as flagged by DAVE). \\

vii) EPIC 211946007.01 is flagged as an eclipsing binary by Gillen et al. (2017) using RV measurements. While DAVE flags this target as a false positive due to odd-even difference, visual inspection did not find the disposition convincing because the transit is not well-sampled due to the very short duration. In addition, there is a field star inside the aperture of EPIC 211946007.01 and it captures the out-of-transit photocenters, but the difference image photocenters lock on the target star itself demonstrating it is the source of the transit signal---which is also corroborated by custom aperture analysis with lightkurve. 

viii) EPIC 211970234.01 is flagged as a false positive by Dressing et al. (2017) because of ``inconsistent transit depth or blended photometry''. DAVE flags this target as a potential false positive because of not transit-like feature but visual inspection marks this as spurious since the duration of the transit is very short and the transits are not well-sampled. In addition, there are multiple field stars in the aperture, and the measured centroid from DAVE is spurious because both the out-of-transit and difference image photocenters lock on the brightest of them instead of on the target star. However, using lightkurve we confirmed that the transit signal is the target star itself and not the field stars. This is shown in Figure \ref{fig:211970234}, where the target's aperture produces a clear transit signal (left panels) but there is no obvious transit signal coming from the field star (right panels). 

ix) Finally, EPIC 228729473.01 is listed as a planet candidate by Mayo et al. (2018) but a false positive due to RV observations by Livingston et al. (2018).

Overall, these considerations demonstrate that barring RV measurements the true disposition of these \OldFPsnotFoundnotFlagged candidates is quite challenging and it is thus not unexpected that they are not flagged as false positives in our catalog. 

\subsubsection{Individual Targets}

\begin{figure*}
\centering
\plotone{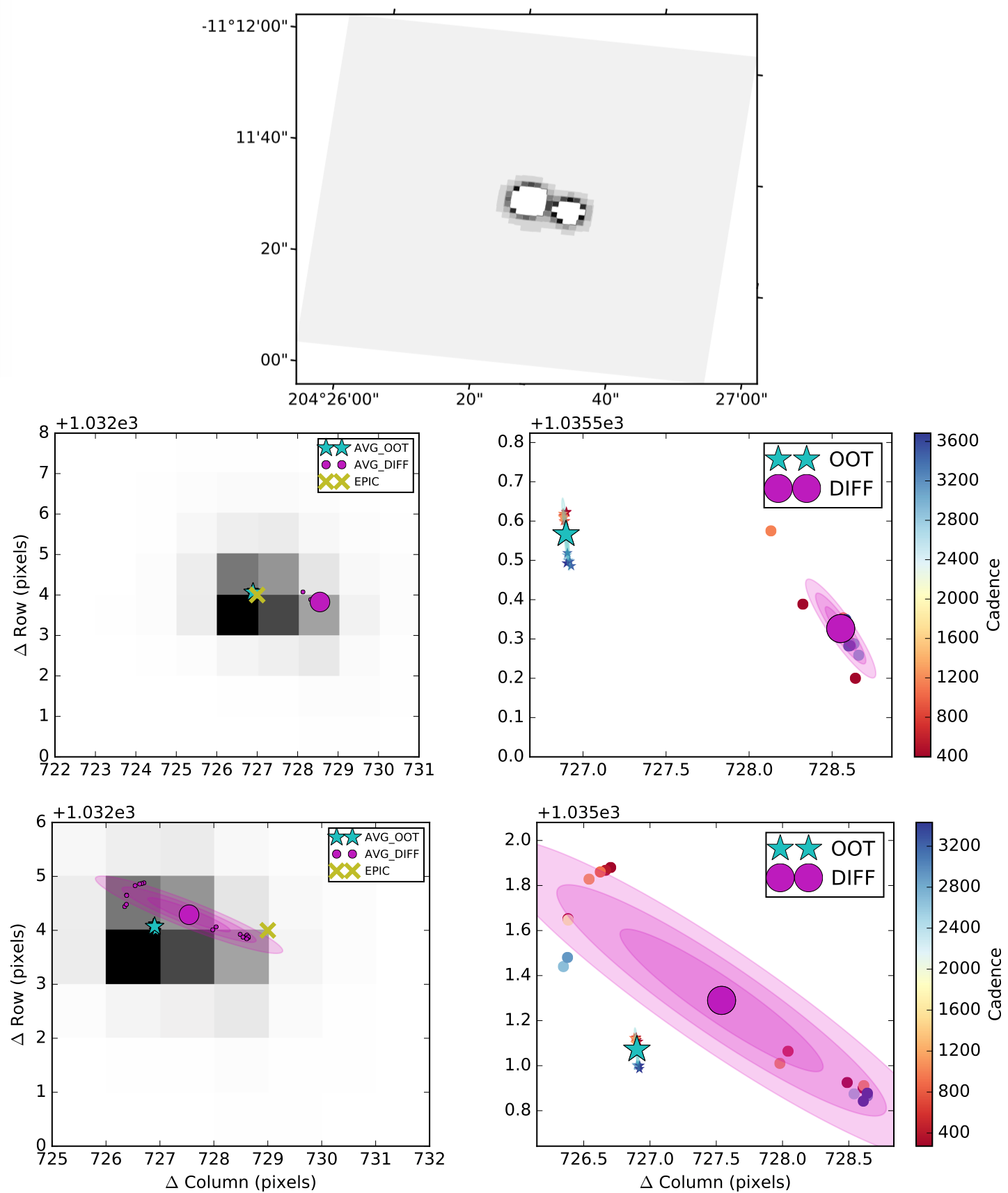}
\caption{Upper panel: 1\arcmin x 1\arcmin~2MASS J-band image of EPIC 212572439.01 (brighter) and EPIC 212572452.01 (fainter). Middle and lower panels: DAVE photocenter measurements for EPIC 212572439.01 (middle panels) and EPIC 212572452.01 (lower panels). The two targets fall in each other's aperture. The pipeline flags both targets as false positives due to measured centroid offset. As discussed in the text, and shown in Figures \ref{fig:212572439_LK} and \ref{fig:212572452_LK}, closer investigation demonstrates that EPIC 212572452.01 is the true source of the transit signal and EPIC 212572439.01 is a false positive.\label{fig:212572439_212572452_COSp}}
\end{figure*}

\begin{figure*}
\centering
\plottwo{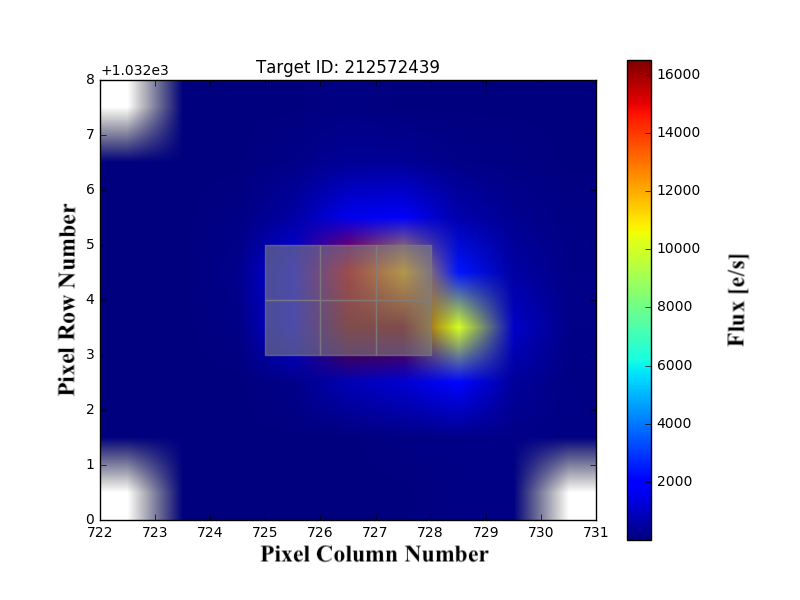}{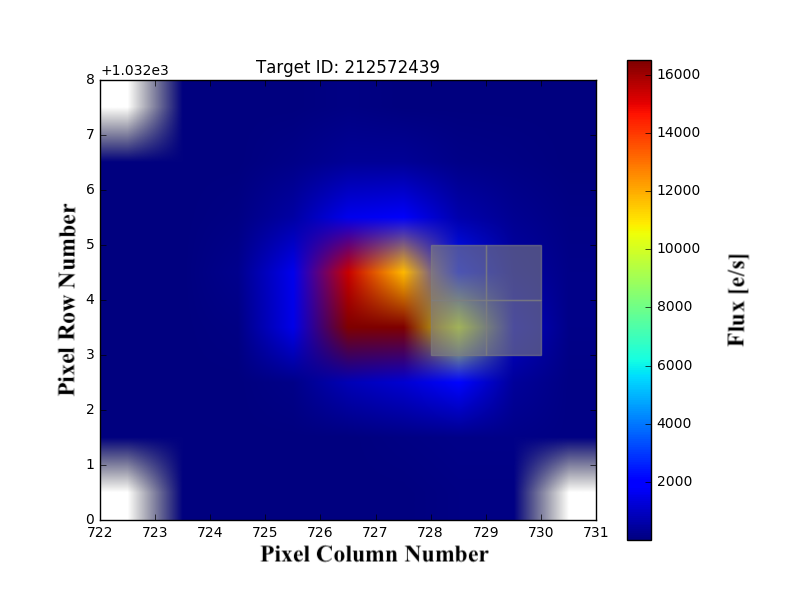}
\plottwo{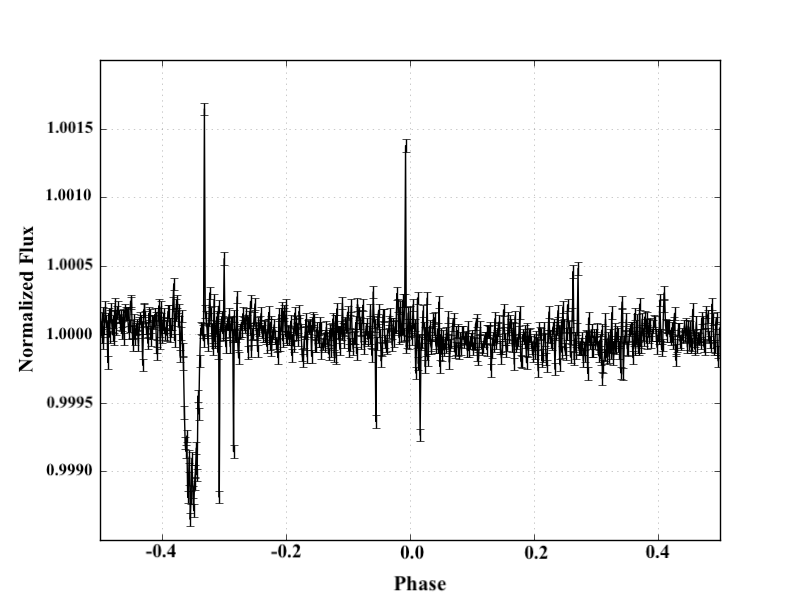}{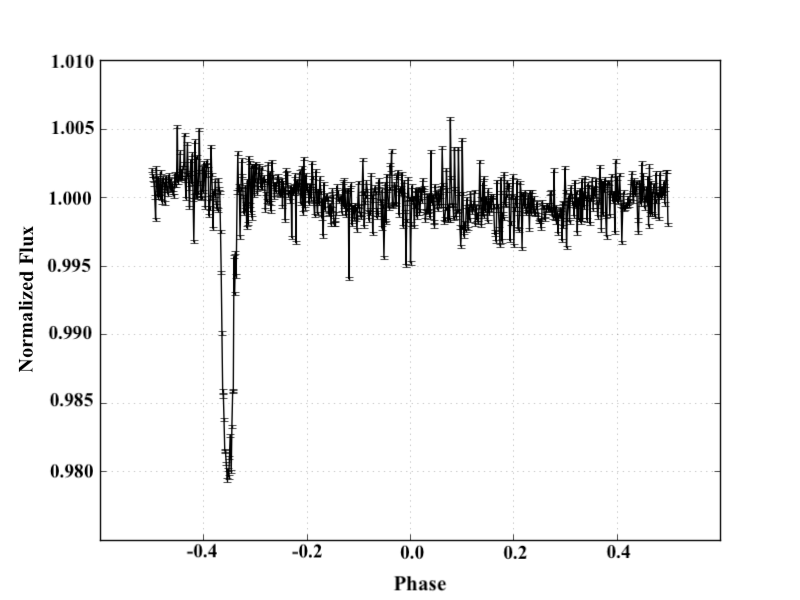}
\caption{Custom apertures (upper panels) and corresponding lightcurves (lower panels) for the false positive EPIC 212572439.01 using lightkurve. The left columns show the results for an aperture centered on the brighter target star (EPIC 212572439.01), and the right columns show the results for an aperture centered on the fainter field star (EPIC 212572452.01). The latter produces a much deeper transit compared to the former, indicating that the field star is the source of the signal.
\label{fig:212572439_LK}}
\end{figure*}

\begin{figure*}
\centering
\plottwo{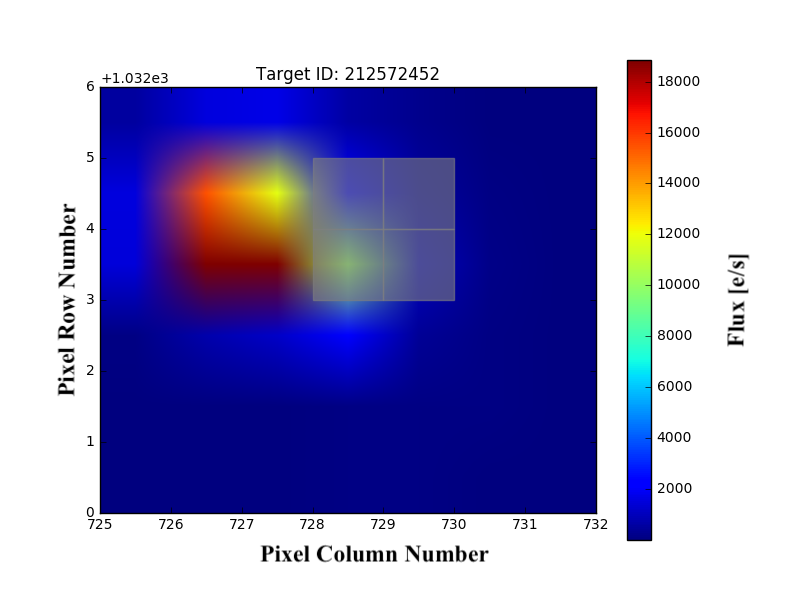}{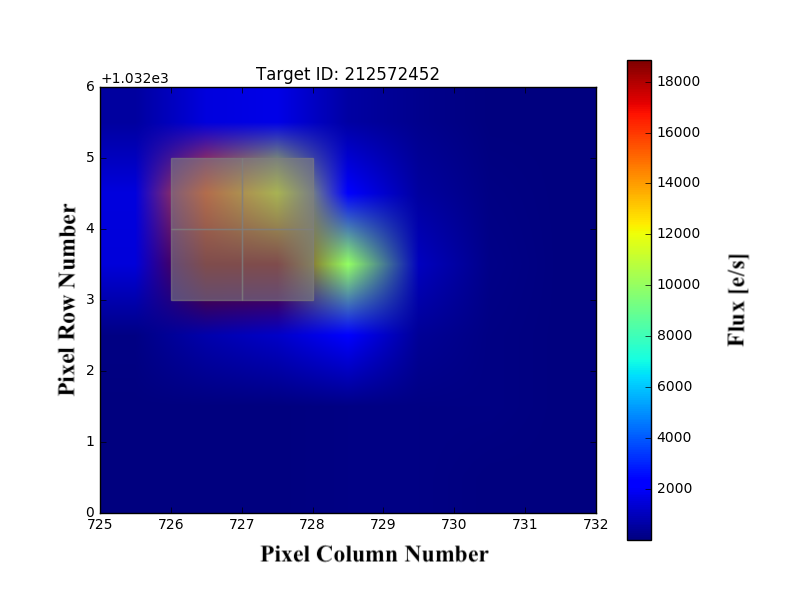}
\plottwo{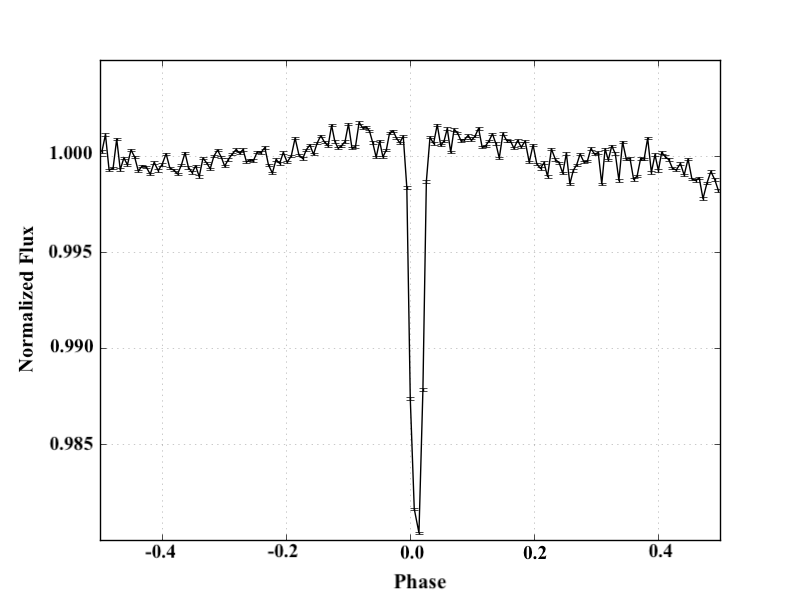}{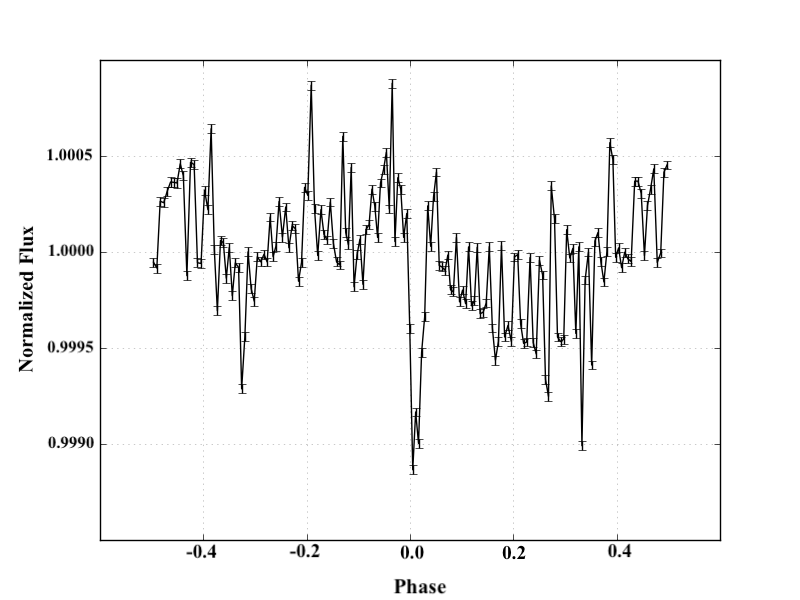}
\caption{Same as Figure \ref{fig:212572439_LK} but for the planet candidate EPIC 212572452.01. Here the target star (EPIC 212572452.01) produces a much deeper transit (left panels) compared to the field star (EPIC 212572439.01, right panels), confirming that the former is the source of the signal. 
\label{fig:212572452_LK}}
\end{figure*}

Targets that happen to fall inside each other's aperture can be particularly challenging to vet. An example is EPIC 212572439.01 (NExScI candidate) and EPIC 212572452.01 (NExScI false positive). Dressing et al. (2017) identify the latter as a false positive due to light contamination from the former. DAVE flags both targets as false positives due to measured significant photocenter shifts during transit. However, closer investigations indicates that EPIC 212572452.01 is the planet candidate and EPIC 212572439.01 is the false positive. Specifically, as shown in Figure \ref{fig:212572439_212572452_COSp} the out-of-transit photocenters for EPIC 212572452.01 lock onto the brighter field star EPIC 212572439.01, while the difference image photocenters lock onto EPIC 212572452.01 itself, indicating it as the true source of the transit signal. In contrast, the out-of-transit photocenters for EPIC 212572439.01 lock onto itself while the difference image photocenters lock onto EPIC 212572452.01, again indicating it as the transit signal. Extracting custom lightcurves with lightkurve confirms this (see Figures \ref{fig:212572439_LK} and \ref{fig:212572452_LK}). 

\subsection{Lessons learned}

During our analysis of the K2 planet candidates, we noticed several issues introduced by the application of an automated vetting pipeline to an inhomogeneous set of lightcurves. Here we list these ``lessons learned''. 

i) In some cases, the center of light can be measured for only one or two cadences at the same roll angle. In such situations DAVE cannot provide a statistically-significant centroid analysis so we used archival images to rule-out potential contamination sources and constrain the offset. 

ii) Sometimes, the test for odd-even difference fails targets with deep transits that after careful visual inspection appear to be bona-fide planet candidates. We realized the importance of taking the systematic red noise (${\rm F_{red}}$) into account when computing the statistical significance of the odd/even metric. To account for this complication, we modified the code to include a comparison between the depth of the transits and the red noise of the lightcurve. For example, we started with automatically failing if ${\rm \sigma_{odd-even} > 4}$, but found it too harsh so compensated with ${\rm (\sigma_{odd-even}/F_{red}) > 4}$, which was more accurate. While this improved the disposition for most of these targets, there were still a few that could not be reconciliated.

iii) To test whether a transit has sufficient signal-to-noise ratio (SNR), DAVE uses a nominal threshold of SNR = 10. However, given that the different lightcurves are processed and detrended by different methods, transits in AGP, EVEREST, and SFF often (but not always) have a different SNR (e.g.~see Figure \ref{fig:agp_vs_everest_}). Thus using a single SNR threshold for all four lightcurve sets is not optimal; for example, we have found that in general a SNR=6-7 works better for PDC. Overall, we note that all automated false position dispositions due to low SNR are marked in our catalog as planet candidates since these candidates may have been discovered in lightcurves customly-detrended beyond what is publicly-available. In addition, if we had K2 transit-injection and recovery experiments, this would help to quantify the detection threshold for each detrending as a function of planet parameters. 

\section{Conclusions}
\label{end}

Capitalizing on our group's unique expertise accumulated as part of Kepler's planet candidate vetting team (e.g.~Mullally et al. 2015; Coughlin et al. 2016), we have developed the fully-automated vetting pipeline DAVE. We have adapted several methods used for vetting Kepler data, i.e. the LPP algorithm (Thompson et al. 2015) and Marshall technique (Mullally et al. 2016) to check if the event is transit-shaped. In addition, we compared the depth to the red noise in the lightcurve, and searched for secondary events. We have also developed a novel method for measuring centroid shifts in the presence of K2's image motion that enable us to measure in-transit image motion at a level approaching that of the Kepler mission.

Using DAVE, we have thoroughly examined \alltargets K2 targets, eliminating a number of different false positives and false alarms, and produced a benchmark catalog of uniformly-vetted planet candidates and false positives. \PCs of these targets, including \NExScIConfirmed confirmed planets, pass our vetting tests and are listed in our catalog as planet candidates. \FPs targets fail one or more of these tests. \OldFPs of these are known false positives, and the rest are new dispositions. The main source of false positives is a significant secondary, non-planetary, eclipse detected in the lightcurve (\FPsSigSec targets), followed by center-of-light offset during transits (\FPsCO targets). A smaller number of targets were dispositioned as false positives due to either out-of-transit modulations (\FPsOOTMOD targets), features that do not appear transit-like (\FPsLCMOD targets), or odd-even difference between consecutive transits (\FPsOE targets). \OldFPsnotFoundbutFlagged targets, while listed on our catalog as planet candidates, are flagged as either having a potentially very deep transit (i.e.~transiting object larger than ${\rm 2R_{Jup}}$, \FPsVDE targets), and/or V-shaped transit signatures (\FPsVSHAPE targets). Two targets that were previously listed as false positives we reclassified as planet candidates (EPIC 211970234.01 and EPIC 212572452.01).

The number of transiting planets is expected to grow significantly in the coming years, especially with the recent launch of the Transiting Exoplanet Survey Satellite (TESS, Ricker et al. 2015) and, looking into the future, the Wide Field Infrared Survey Telescope (WFIRST, Spergel et al. 2015) as well. The knowledge gained here will be key when applying DAVE and other vetting diagnostic tools to the large number of planet candidates expected to be found in a multitude of community-produced lightcurves from the TESS pixel level data and full frame images. TESS will produce full-frame images at the same cadence as K2, and these will contain ${\sim 25-30 million}$ persistent light sources---amongst which there will be thousand of transiting planet candidates (Barclay et al. 2018). Quickly vetting these with DAVE will be critical for the rapid follow-up needed to obtain mass measurements, for comparing the community lightcurves, and for the development of a uniform catalog of TESS planets. We are currently developing DAVE to be directly applicable to TESS data, and are already testing it on recently released planet candidate from TESS. In addition, planet validation will help enable prioritization of the best targets for atmospheric characterization with the Hubble Space Telescope and JWST.

The DAVE catalog will be hosted, in a table format, by the Mikulski Archive for Space Telescopes (MAST) and will be properly maintained, archived, and documented at the archive in perpetuity. We will incorporate any additional candidates found by the community so that all planetary candidates found in K2 data will have a consistent set of statistics and vetting products available. In addition, we hope to continue improving our algorithms, refactor our code to improve readability, and add documentation to lower the barrier to community use, while maintaining our open access policy. 

\clearpage

%
%
%
%
%
%
\clearpage
\begin{table*}[ht]
\begin{center}
\footnotesize
\scriptsize
\caption{Number of candidates with significant transits in each dataset.
\label{tab:RMN_}}
\begin{tabular}{ccccc}
\hline
\hline
Detrending & AGP & EVE & PDC & SFF \\
\hline
Significant Transits (``S'')$^\dagger$ & 541 & 693 & 537 & 719 \\
Transits not significant (``N'')$^\ddagger$ & 63 & 65 & 223 & 45 \\
N/A$^\S$ & 168 & 14 & 12 & 8 \\
Total number & 772 & 772 & 772 & 772 \\
\hline
Fraction significant (S / (S + N)) & 0.9 & 0.91 & 0.71 & 0.94 \\
\hline
\hline
\multicolumn{5}{l}{$^\dagger$: ${\rm SNR\sim5-10}$, depending on the detrending }\\
\multicolumn{5}{l}{$^\ddagger$: ${\rm SNR<5}$, depending on the detrending }\\
\multicolumn{5}{l}{$^\S$: Data not available }\\\end{tabular}
\\
\end{center}
\end{table*}

\begin{sidewaystable*}[htbp]
\centering
\tiny
\caption{DAVE disposition of confirmed planets listed on NExScI as of August, 6, 2018. The corresponding columns are: EPIC ID, Period, Epoch, automated false positive reason for AGP, EVEREST, PDC and SFF lightcurves respectively, Final disposition in DAVE catalog, False positive reason, NExScI disposition, Comments. The corresponding abbreviations are listed in Table \ref{tab:abbreviations}. The lower part of the table shows the 4 targets that are listed in NExScI as confirmed planets---and are marked as such in the DAVE catalog---but may be false positives due to potential centroid offset. This table and the next are available in their entirety as a single table in electronic format. 
\label{tab:CP}}
\begin{tabularx}{\linewidth}{cccccccccccX}
\hline
\hline
EPIC ID & Period & Epoch & AGP Disp & EVE Disp & PDC Disp & SFF Disp & Final Disp & FP Reason & NExScI Disp & Comments & DAVE URL\\
\hline
201110617 & 0.813 & 2787.557292 & Low SNR & None &  Low SNR & None & PC & - & CP & & http://...\\
201130233 & 0.365 & 2788.542394 & SS & N/A & N/A & SS & PC & - & CP & SS NC & ...\\
201155177 & 6.687 & 2015.11927 & N/A & None & Low SNR & None & PC & - & CP & ... & ... \\
201166680 & 18.105 & 2783.302493 & CO & None & Low SNR & None & PC & - & CP & AGP~COSp & ... \\
... & ... & ... &... & ... &... & ... &... & ... &... & ... & ... \\
\hline
201211526 & 21.07 & 2797.621875 & CO & CO & CO & CO & PC & - & CP & pCO & ... \\
201629650 & 40.063 & 2019.587878 &... & ... &... & ... & PC & - & CP & pCO, FSAp & ... \\
206119924 & 4.656 & 2179.540469 & CO & Low SNR & Low SNR & LCMOD & PC & - & CP & pCO, FSAp & ...\\
211594205 & 16.994 & 2349.48838 &... & ... &... & ... & PC & - & CP & pCO & ...\\
\hline
\hline
\end{tabularx}
\end{sidewaystable*}

\begin{sidewaystable*}[htbp]
\centering
\tiny
\caption{Same as Table \ref{tab:CP} but for the false positives. Upper third of the table: False positives from DAVE and NExScI; Middle third of the table: Planet candidates from DAVE but false positives from NExScI; Lower third of the table: False positives from DAVE but planet candidates from NExScI. See Table \ref{tab:abbreviations} for abbreviations.
\label{tab:FP}}
\begin{tabularx}{\linewidth}{cccccccccccccX}
\hline
\hline
EPIC ID & Period & Epoch & AGP Disp & EVE Disp & PDC Disp & SFF Disp & Final Disp & FP Reason & NExScI Disp & Comments & DAVE URL\\
\hline
210954046 & 0.95 & 2263.967875 & CO & None &  None & CO & FP & CO & FP (RV) & & ... \\
206155547 & 24.387 & 2177.271763 & SS & SS & None & SS & FP & SS & FP (RV) & No~centroid & ...\\
210401157 & 1.316 & 2264.541126 & SS & SS & SS & SS & FP & OOTMOD & FP$^\ddagger$ & pSS & ...\\
... & ... &... &... & ... &... & ... &... & ... &... & ... & ... \\
\hline
210754505 & 0.807 & 2264.106897 & None & OE & None & None & PC & - & FP (OOTMOD) & pOE,~pOOTMOD & ...\\
210414957 & 0.97 & 2264.112891 &None & None &None &None & PC & - & FP (OOTMOD) & pOOTMOD & ...\\
206065006 & 25.277 & 2186.653484	 &Low~SNR & Low~SNR & Low~SNR & Low~SNR & PC & - & FP (Vespa) & Low SNR& ...\\
... & ... &... & ... & ... &... & ... &... & ... &... & ... & ... \\
\hline
212572439 & 2.582 & 2423.58774 & CO & CO & CO & CO & FP$^\ddagger$$^\ddagger$ & CO & PC & FSApST & ...\\
... & ... &... & ... &... &... & ... &... & ... &... & ... & ... \\
\hline
\hline
\multicolumn{13}{l}{$^\ddagger$: ``strange, composite spectrum'' according to Crossfield et al. (2016)} \\
\multicolumn{13}{l}{$^\ddagger$$^\ddagger$: Same ephemeris as EPIC 212572452, which is the true source of the transit signal.} \\
\end{tabularx}
\end{sidewaystable*}

\begin{table*}[ht]
\begin{center}
\footnotesize
\scriptsize
\caption{False positives listed in the DAVE catalog (http://keplertcert.seti.org/DAVE/). See Table \ref{tab:abbreviations} for abbreviations.
\label{tab:FP_final}}
\begin{tabular}{l|l|l}
\hline
\hline
Disposition Flag & Reason for disposition & Number of targets \\
\hline
SS & Significant Secondary & \FPsSigSec\\
CO & Photocenter shift during transit & \FPsCO\\
OOTMOD & Out-of-transit modulations in-phase with the transits & \FPsOOTMOD \\
LCMOD & Feature does not appear transit-like & \FPsLCMOD \\
OE & Transit depth alternates between consecutive transits& \FPsOE \\
\hline
Total & & \FPs$^\dagger$$^\ddagger$ \\
\hline
\hline
\multicolumn{3}{l}{$^\dagger$: \FPsMoreThanOne targets exhibit more than one false positive indicator.}\\
\multicolumn{3}{l}{$^\ddagger$: All low-SNR dispositions are marked as planet candidates in the DAVE catalog by default. See text for details.} \\
\end{tabular}
\\
\end{center}
\end{table*}

\begin{table*}[ht]
\begin{center}
\footnotesize
\scriptsize
\caption{Abbreviations.
\label{tab:abbreviations}}
\begin{tabular}{cc}
\hline
\hline
Abbreviation & Description \\
\hline
AGP & AGP lightcurves (Aigrain et al. 2015) \\
CO & Centroid Offset \\
COSp & Centroid Offset Spurious \\
CP & Confirmed Planet \\
EVEREST & EVEREST lightcurves (Luger et al. 2016, 2018) \\
FP & False Positive \\
FSAp & Field Star(s) in Aperture \\
FSApST & Field Star(s) in Aperture, the Source of the Transit \\
LCMOD & Lightcurve Modulations \\
N/A & Disposition and/or lightcurve Not Available \\
NC & Not Convincing \\
OE & Odd-Even difference \\
OOTMOD & Out-of-transit Modulations \\
PC & Planet Candidate \\
pCO & Potential Centroid Offset \\
PDC & PDC lightcurves (Smith et al. 2012) \\
pFP & Potential False Positive \\
pOOTMOD & Potential Out-of-Transit Modulations \\
pSS & Potential Significant Secondary \\
RFS & Recommend Further Scrutiny \\
SFF & SFF lightcurves (Vanderburg et al. 2014) \\
SNR & Signal-to-noise ratio \\
TESS & Transiting Exoplanet Survey Satellite \\
TLM & Transit-Like Metric \\
pVDE & Potentially Very Deep Eclipse \\
VSHAPE & V-shape transit \\
\hline
\hline
\end{tabular}
\\
\end{center}
\end{table*}

\clearpage
\acknowledgments
We thank the referee for the insightful comments that helped us improve this manuscript. This paper includes data collected by the K2 mission. Funding for the K2 mission is provided by the NASA Science Mission directorate. The data presented in this paper were obtained from the Mikulski Archive for Space Telescopes (MAST), operated by the Association of Universities for Research in Astronomy, Inc., under NASA contract NAS5-26555. The NASA Exoplanet Archive is operated by the California Institute of Technology, under contract with NASA under the Exoplanet Exploration Program. VK, EQ and JC gratefully acknowledge support from NASA via grant NNX17AF81G. SM gratefully acknowledges support from NASA via grant NNX16AE74G. FM gratefully acknowledges support from NASA via grant NNX16AJ19G.  

{\it Software}: DAVE (https://github.com/exoplanetvetting/DAVE), Kepler Science Data Processing Pipeline (https://github.com/nasa/kepler-pipeline), Centroid Robovetter (Mullally 2017), LPP Metric (Thompson et al. 2015), Scipy package (https://www. scipy.org)

\bibliography{dave}

{\bf Bibliography}

{\noindent Adams. E. R., Jackson, B. \& Endl, M. 2016, \aj, 152, 47 \\}
Aigrain, S., Hodgkin, S. T., Irwin, M. J., Lewis, J. R., \& Roberts, S. J. 2015, MNRAS, 447, 2880 \\
Ansdell, M., Ioannou, Y., Osborn, H. P. et al. 2018, \apj, 869, 7 \\
Baglin, A., et al. 2006, 36th COSPAR Scientific Assembly, 36, 3749 \\
Barclay, T.; Pepper, J.; \& Quintana, E. V. 2018, 2018arXiv180405050B \\
Barros, S. C. C.; Demangeon, O.; \& Deleuil, M. 2016, \aap, 594, 100 \\
Batalha, N. M., Borucki, W. J., Bryson, S. T., et al. 2011, \apj, 729, 27 \\
Borucki, W. J.; Koch, D.; Basri, G. et al. 2010, Science, 327, 977 \\
Borucki, W. J.; Koch, D. G.; Basri, G. et al. 2011, \apj, 736, 19 \\
Bryson, S. T., Jenkins, J. M., Klaus, T. C., et al. 2010, in Society of Photo-Optical Instrumentation Engineers (SPIE) Conference Series, Vol. 7740 \\
Bryson, S. T., Jenkins, J. M., Gilliland, R. L., et al. 2013, PASP, 125, 889 \\
Burke, C. J.; Christiansen, J. L.; Mullally, F. et al. 2015, \apj, 809, 8 \\
Cabrera, J., Barros, S. C. C., Armstrong, D., et al. 2017, \aap, 606, A75 \\
Catanzarite, J. H. 2015, Autovetter Planet Candidate Catalog for Q1-Q17 Data Release 24 \\
Christiansen, J. L., Vanderburg, A., Burt, J., et al. 2017, \aj, 154, 122 \\
Coughlin, J. L., Thompson, S. E., Bryson, S. T., et al. 2014, \aj, 147, 119 \\
Coughlin, J. L., Mullaly, F., Thompson, S. E., et al. 2016, \apjs, 224, 12 \\
Crossfield, I. J. M., Ciardi, D. R., Petigura, E. A., et al. 2016, \apjs, 226, 7 \\
Crossfield, I. J. M., Guerrero, N., David, T. et al. 2018, 2018arXiv180603127C \\
Deming, D., Knutson, H.. Kammer, J. et al. 2015, \apj, 805, 132 \\
Dressing, C. D. \& Charbonneau, D. 2015, \apj, 807, 45 \\
Dressing, C. D., Vanderburg, A., Schlieder, J. E., et al. 2017, \aj, 154, 207 \\
Gillen, E., Hillenbrand, L. A., David, T. J. et al. 2017, \apj, 849, 11 \\
He, X., \& Niyogi, P. 2004, Advances in Neural Inoformation Processing Systems, 16, 37 \\
Howell, Steve B.; Sobeck, Charlie; Haas, Michael; 2014, PASP, 126, 398H \\
Jenkins, J. M., (ed.) 2017, Kepler Data Processing Handbook (KSCI-19081-002) \\
Kaltenegger \& Traub 2009, ApJ, 698, 519 \\
Livingston. J. H., Endl, M., Dai. F. et al. 2018, \aj, 156, 78 \\
Luger, R., Agol, E., Kruse, E., et al. 2016, \aj, 152, 100 2016 \\
Luger, R., Kruse, E., Foreman-Mackey, D. et al. 2018, \aj, 156, 99 \\
Matijevic, G. Prsa, A., Orosz, J.A. et al. 2012, \aj, 143, 123 \\
Mayo, A. W., Vanderburg, A., Latham, D. W., et al. 2018, \aj, 155, 136 \\
Montet, B. T., Morton, T. D., Foreman-Mackey, D., et al. 2015, \apj, 809, 25 \\
Morton, T. 2012, \apj, 761, 6 \\
Mullally, F., Coughlin, J. L., Thompson, S. E., et al. 2015, \apjs, 217, 31 \\
Mullally, F., Coughlin, J. L., Thompson, S. E., et al. 2016, PASP, 128, 074502 \\
Mullally, F. 2017, Planet Detection Metrics: Automatic Detection of Background Objects Using the Centroid Robovetter (KSCI-19115-001) \\
Petigura, E. A., Howard, A. W., Marcy, G. W., et al. 2017, \aj, 154, 107 \\
Pope, B. J. S., Parviainen, H., \& Aigrain, S. 2016, MNRAS, 461, 3399 \\
Ricker, G. R., Winn, J. N., Vanderspek, R., et al. 2015, Journal of Astronomical
Telescopes, Instruments, and Systems, 1, 014003 \\Rizzuto, A. C., Mann, A. W., Vanderburg, A. et al. 2017, \aj, 154, 224 \\
Rowe, J. F., Coughlin, J. L., Antoci, V., et al. 2015, \apjs, 217, 16 \\
Shallue, C. J., \& Vanderburg, A. 2018, \aj, 155, 94 \\
Shporer, A., Zhou, G., Vanderburg, A., et al. 2017, \apj, 847, 18 \\
Smith, J. C., Stumpe, M. C., Van Cleve, J. E. et al. 2012, PASP, 124, 1000 \\
Spergel, D. Gehrels, N., Baltay, C. et al. 2015, 2015arXiv150303757S \\
Stassun, K. G., Oelkers, R. J., Pepper, J. et al. 2018, \aj, 156, 183 \\
Thompson, S. E., Mullally, F., Coughlin, J. L., et al. 2015, ApJ, 812, 46 \\
Thompson, S. E., Caldwell, D. A., Jenkins, J. M., et al. 2016, Kepler Data Release 25 Notes (KSCI-19065-002) \\
Thompson, S. E.; Coughlin, J. L.; Hoffman, K. et al. 2018, \apjs, 235, 38\\
Vanderburg, A. \& Johnson, J. A. 2014, PASP, 126, 948 \\
Vanderburg, A., Latham, D. W., Buchhave, L. A., et al. 2016, \apjs, 222, 14 \\
Vincius, Z., Barentsen, G., Hedges, C. et al. 2018, KeplerGO/lightkurve, 10.5281/zenodo.1181928 \\

\end{document}